\documentclass[aip,reprint,superscriptaddress,amsmath,amssymb,amsfonts,floatfix,showkeys,showpacs]{revtex4-2}

\usepackage{dcolumn}  
\usepackage{multirow }
\usepackage{academicons}
\usepackage{soul}
\usepackage{hyperref}

\usepackage[T1]{fontenc}
\usepackage[utf8]{inputenc}

\usepackage{verbatim}
\usepackage{cprotect}

\usepackage{bm} 
\usepackage{hyperref}
\usepackage{natbib}
\usepackage{physics}
\usepackage[english]{babel}
\usepackage{babelbib}
\usepackage{graphicx}
\usepackage{epstopdf}
\usepackage{ifpdf} 
\usepackage{epsfig}
\usepackage{color}
\usepackage[usenames,dvipsnames]{xcolor}
\usepackage{subfigure}
\usepackage{amsmath}

\usepackage{amsmath} 
\usepackage{amssymb} 
\usepackage{mathrsfs} 

\usepackage{subfigure}

\newcommand{\rvsd}[1]{\textcolor{black}{#1}}

\newcommand{\orcid}[1]{\href{https://orcid.org/#1}{\includegraphics[height = 8pt]{orcid_logo.png}}}
\usepackage{hyperref}
\usepackage{academicons}
\usepackage{xcolor}

\begin{document}

\affiliation{Department of Physics, College of Science, Sultan Qaboos University, P.O. Box 36, Muscat 123, Oman}
\affiliation{School of Nano Science, Institute for Research in Fundamental Sciences (IPM), Tehran 19538-33511, Iran}
	
\title{Smart navigation through a rotating barrier: Deep reinforcement learning with application to size-based separation of active microagents}

\author{Mohammad Hossein Masoudi} 
	\affiliation{School of Nano Science, Institute for Research in Fundamental Sciences (IPM), Tehran 19538-33511, Iran}
\author{Ali Naji} 
\thanks{Author to whom correspondence should be addressed: \texttt{a.naji1@squ.edu.om}.}
\affiliation{Department of Physics, College of Science, Sultan Qaboos University, P.O. Box 36, Muscat 123, Oman}
\affiliation{School of Nano Science, Institute for Research in Fundamental Sciences (IPM), Tehran 19538-33511, Iran}
    
\begin{abstract}
\rvsd{We employ deep reinforcement learning methods to investigate shortest-time navigation strategies for smart active Brownian particles (microagents), which self-propel through a rotating potential barrier in a static, viscous, fluid background. The microagent's motion begins at a specified origin and terminates at a designated destination. The potential barrier is modeled as a localized, repulsive Gaussian potential with finite support, whose peak location rotates at a given angular velocity about a fixed center within the plane of motion. We use the Advantage Actor-Critic approach to train microagents for their origin-to-destination navigation through the barrier. By employing this approach, we demonstrate that the rotating potential (as opposed to a static one) enables size-based sorting and separation of the microagents. In other words, microagents of different radii arrive at the destination at sufficiently well-separated average times, facilitating their  sorting. The efficiency of particle sorting is quantified  by introducing specific separation measures. We also demonstrate how training the microagents in a noisy background, as opposed to a noise-free one, can improve the precision of their size-based sorting. Our findings suggest promising avenues for future research on smart active particles equipped with deep reinforcement learning to navigate complex environments, particularly in microscale applications.}
\end{abstract}

\maketitle

\section{Introduction}\label{sec:intro}

\rvsd{Active matter refers to a broad class of soft materials composed of active particles that  convert ambient free energy (often from a surrounding fluid) into self-propulsive motion \cite{marchetti2013hydrodynamics}.} Examples of active particles include molecular motors in living cells, microorganisms such as bacteria, and synthetic colloids such as Janus particles \cite{elgeti2015physics,bechinger2016ap,paxton2004catalytic}. These microagents display nearly ballistic self-propulsive motion over a characteristic run time, which is followed or accompanied by stochastic reorientation of their self-propulsion axis. Systems of active particles exhibit a rich phenomenology due to their inherent nonequilibrium dynamics which \rvsd{combines with} other factors (such as noncentral interparticle alignment effects, long-range hydrodynamic interactions,  strong particle coupling to confining boundaries, etc) \rvsd{to produce a wide range of complex collective behaviors}  \cite{vicsek2010coll,cates2014mips,solon2015pressure,chat2008coll,ginelli2010large,zottl2016emergent, lauga2016bacterial,toner2005hydrodynamics,buttinoni2013dynamical,zottl2012nonlinear,son2015live,lauga2016bacterial,yan2016reconfiguring,redner2013structure, theurkauff2012dynamic,romanczuk2012ABP}. 

The mechanisms of self-propulsion can vary widely. Biological microorganisms propel by modulating various modes of motion in their cellular organelles such as flagella and cilia \cite{Gib1982cilia, Berg2003,lauga2016bacterial}. Synthetic active colloids  \cite{Howse2007,zottl2016emergent, paxton2004catalytic,rao2018self,poggi2017janus,bar2020self}, on the other hand, utilize catalytic surface activity to produce self-propulsive force, driving their autonomous locomotion in fluid media. In most cases, the pattern of motion may vary also in response to surrounding stimuli and external fields, including chemical nutrient fields (as in chemotactic bacteria \cite{berg1977escherichia}), light intensity (as in phototactic active particles  \cite{palacci2013living}),  Earth's gravitational field  (as in gyrotactic algae  \cite{goldstein2015green}), and external magnetic fields (as in magnetotactic bacteria  \cite{blakemore1975magnetotactic,erglis2007mag} and active dipolar spheroids  \cite{shabanniya2020active}).  
 
Despite the diversity of examples, the generic aspects of self-propelled motion in active microagents can be captured by minimal models, such as the active Brownian particle (ABP) model \cite{MARCHETTI201634}. In its simplest form, the ABP model assumes a constant self-propulsion speed and a reorientation mechanism driven by the rotational Brownian noise. The simplicity of this and other similar models has enabled their integration with reinforcement learning (RL) methods to study optimal navigation (or other task performances) by self-propelled agents in  complex environments \cite{stark2019opt,colabrese2017flow,Gustavsson2017,biferale2019zermelo,nasiri2022reinforcement,gunnarson2021learn,buzzicotti2021opt,zou2022gait,Borra2022RL,muinos2021reinforcement,Putzke2023}. Recent works in this context include path planning of ABPs \cite{stark2019opt}  to discover the optimal shortest-time trajectories through a static Mexican-hat potential \cite{stark2019opt}, and planning optimal escape strategies for gyrotactic particles trapped in the Green-Vortex flow \cite{colabrese2017flow}. RL methods have also been utilized for active navigation through  \rvsd{chaotic and  turbulent flows \cite{Gustavsson2017,buzzicotti2021opt,Calascibetta2023}. On the other hand, the advancement of deep reinforcement learning (DRL) techniques in obtaining nonlinear optimal policy functions through neural networks has proven advantageous in developing enhanced strategies for active navigation through unsteady vortical flows \cite{gunnarson2021flow},  shear flows through a pipe and Gaussian random potentials \cite{nasiri2022reinforcement}.}

Despite the aforementioned advances in studying smart navigation through {\em global} time-varying flows \cite{gunnarson2021flow,buzzicotti2021opt}, navigation of active agents through {\em localized}, time-varying perturbations, specifically a moving barrier in an otherwise static  environment, remains largely unexplored. \rvsd{This is the scenario we aim to study in this work using a simple two-dimensional (2D) model, in which the barrier is modeled as a repulsive Gaussian potential with finite support (range of action), whose peak location rotates at a fixed angular velocity around a specified center within the plane of motion. For the agent, on the other hand, we introduce a smart active Brownian particle (sABP) model; that is, a self-propelled Brownian particle equipped with a DRL algorithm. Specifically, the agent is trained using the Advantage Actor-Critic (A2C) approach to navigate optimal, shortest-time trajectories from a specified origin to a designated destination, encountering the barrier along its path.}

\rvsd{For the specific case of interest in this work, the environment being navigated by the agent exhibits a regular pattern of temporal perturbations due to the periodic motion of the barrier. This is expected to enhance controllability of the shortest-time navigation process in terms of agent-specific parameters.} We demonstrate that our model can indeed be used to sort and separate microparticles based on their size, a process which is of major relevance to microfluidic applications \cite{Son2015,Shields2015,Lenshof2010}. When the barrier is static (non-rotating), the size-based separation is found to be absent. We also discuss how the rotational Brownian noise can be integrated into agent-training procedures via the proposed DRL algorithm and, hence, show how the presence of noise affects the predicted size-based sorting of the agents within our model. \rvsd{Interestingly, we find that training the agent in a noisy background enhances the precision of size-based sorting compared to training it in a noise-free environment (note that the sorting takes place in the simulation stage, following a training stage where the optimal navigation policy for the agent   is obtained through an adequate number of  training episodes, as will be detailed later.)}

\rvsd{The paper is organized as follows: Our model and methods are introduced in Secs. \ref{sec:model} and \ref{sec:methods}. The results for optimal trajectories obtained from training in a noise-free background are presented in Secs. \ref{subsec:trajectories} and \ref{subsec:sizedet}. In Secs. \ref{subsec:sizedetthermal} and \ref{subsec:noisetrain}, we discuss the impact of training in a noisy background on our results. The paper is concluded  in Sec. \ref{sec:conclusions}.}

\begin{figure}[t]
\centering
\includegraphics[width = 8.5cm]{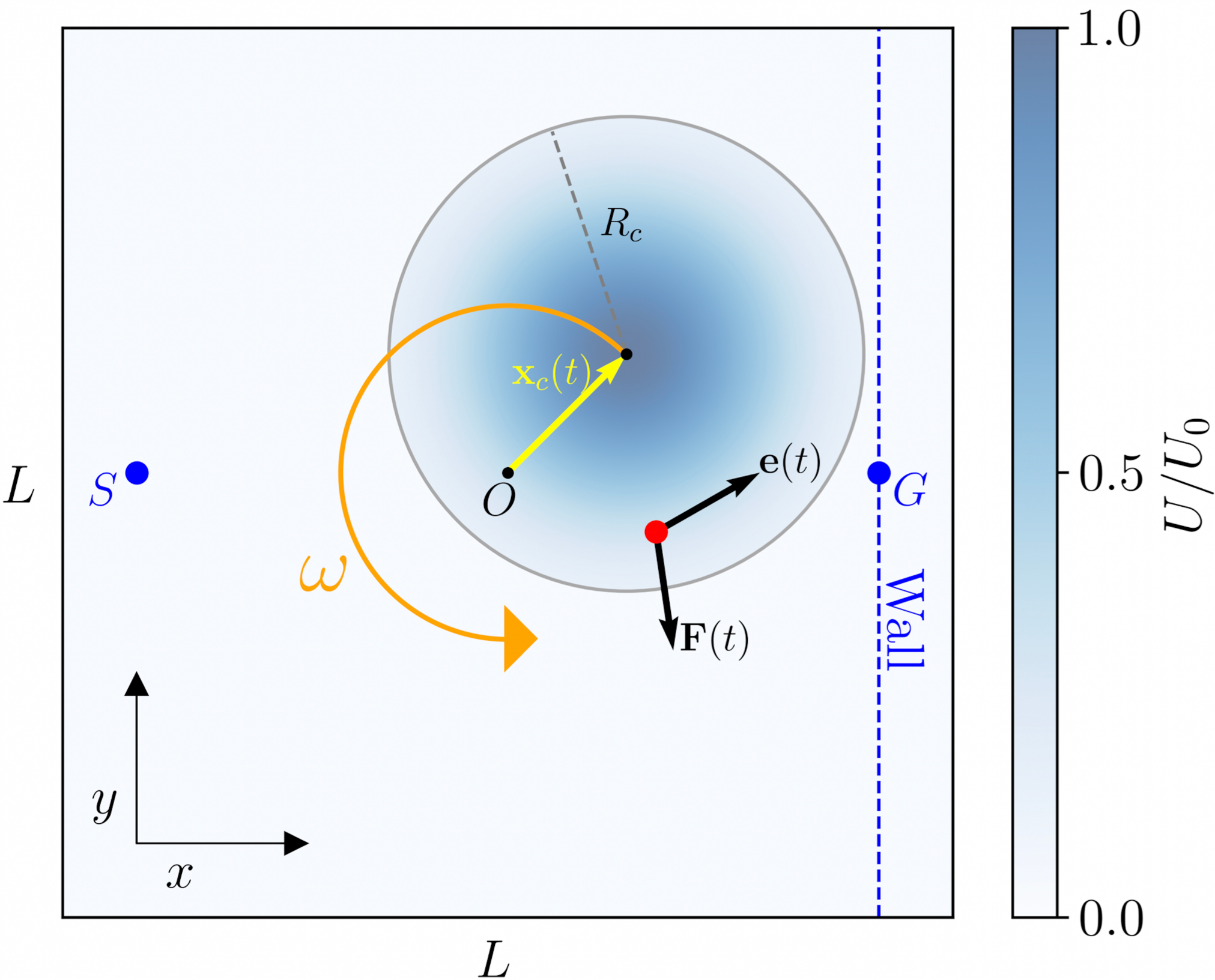}
\caption{\rvsd{An active microagent (red dot), with the orientation unit vector $\mathbf{e}(t)$, navigates shortest-time trajectories from S to G on a 2D plane. On its path, the microagent encounters a repulsive Gaussian potential barrier (color-coded disk-shaped region) peaked at $\mathbf{x}_c(t)$. The barrier peak rotates at a constant angular velocity $\omega$ (orange arrow) around the fixed point O, which is taken as the midpoint between S and G. The vertical dashed line shows the target wall. The system is confined within a square box of side length $L$, which is also centered at O. The range of the potential is marked by the dark gray circle, and the corresponding force on the microagent  is shown by $\mathbf{F}(t)$. The colorbar shows the rescaled potential strength, $U / U_0$ (see the text).}}
\label{Gaus_rot}
\end{figure}

\section{Model}
\label{sec:model}

Our model comprises three main components: (i) a microagent modeled as a colloidal particle that, in the absence of learning strategies, exhibits a self-propelled Brownian motion, \rvsd{in accordance with the well-established models of minimal active Brownian particles (ABPs) \cite{MARCHETTI201634}}; (ii) an inhomogeneous environment where the microagent encounters a finite-range, dynamically evolving potential barrier as it moves from a prescribed  start point (S) to a designated destination (goal G) within the $x-y$ plane (see Fig. \ref{Gaus_rot}); and (iii) a DRL framework that trains the model microagent to achieve shortest-time navigation through the  environment. 

\rvsd{The baseline Brownian motion in (i) follows a standard, overdamped Langevin equation involving viscous (Stokes) drag  and thermal noise terms. These features arise from an implicitly assumed fluid in the background, and can thus be regarded as parts of the environment (ii) as well. We may refer to the microagent (or, interchangeably, the agent) equipped with the learning component (iii) also as a {\em smart active Brownian particle (sABP)}, by analogy with {\em nonsmart} ABPs mentioned above \cite{MARCHETTI201634}. The detailed aspects of  components (i)-(iii) will be discussed  in what follows.}

\subsection{Equations of motion without learning}\label{Eqwithoutlearning}

\rvsd{The model sABP is assumed to be a spherical particle of radius $a_p$ with position and orientation coordinates ${\mathbf{x}}(t)=(x(t),y(t))$ and $\mathbf{e}(t)=(\cos{\theta(t)},\sin{\theta(t)})$, respectively. It is confined to a 2D square box of side length $L$, centered at  $O(0,0)$ (not to be confused with the start point, S). We shall also refer to the vertical dashed line drawn through the destination point G in Fig. \ref{Gaus_rot} as the target wall (or, the wall, for brevity). The particle trajectories in the simulations terminate upon hitting the wall, as we shall clarify later.}

In the absence of learning, ${\mathbf{x}}(t)$ and the angular coordinate $\theta(t)$, which is measured  relative to the $x$-axis, evolve according to the Langevin equations \cite{MARCHETTI201634} (modeling component i)
\begin{eqnarray}
	\label{vel_tr}
&&    \dot{\mathbf{x}}(t) = v_0 \mathbf{e}(t) + \gamma_T^{-1}\mathbf{F}(\mathbf{x}(t),t),\\
    \label{vel_rot}
&&     \dot{\theta}(t) = -\alpha\, \Omega(\mathbf{x}(t),\theta(t),t)+{\xi}_R(t), 
\end{eqnarray}
\rvsd{where $v_0$ is the constant self-propulsion velocity along the unit vector   $\mathbf{e}(t)$, $\gamma_T$ is the translational friction coefficient, and $\Omega(\mathbf{x}(t),\theta(t),t)$ is a non-dimensional deterministic angular velocity associated with a realignment rate (or strength)  $\alpha$, as will be clarified later.} 

\rvsd{Furthermore, $\mathbf{F}(\mathbf{x}(t), t) = -\nabla U(\mathbf{x}(t), t)$ (which we may denote later by $\mathbf{F}(t)$ or $\mathbf{F}$ for simplicity) is the deterministic force on the microagent due to its interaction with the repulsive potential barrier $U(\mathbf{x}(t), t)$  (modeling component ii). The barrier is assumed to be a rotating Gaussian with peak location at $\mathbf{x}_c(t)=|\mathbf{x}_c(t)|(\cos \omega t,\sin \omega t)$, standard deviation $\sigma$, and range (cutoff radius) $R_c$; i.e.,}  
\begin{equation}
U(\mathbf{x}(t),t) = U_0 \,\mathrm{e}^{-\lvert\mathbf{x}(t) - \mathbf{x}_c(t)\rvert^2/{2\sigma^2}}
\label{rot_potential}
\end{equation}
when $\lvert\mathbf{x}(t)-\mathbf{x}_c(t)\rvert \leq R_c$, and $U(\mathbf{x}(t),t)=0$ otherwise. \rvsd{The barrier rotates around the center, O, at constant angular velocity $\omega$. For simplicity, we set $\sigma=R_c/2$.}
 
Additionally, in Eq. \eqref{vel_rot}, ${\xi}_R(t)$ is the rotational Brownian noise with zero mean and two-point correlator $\langle  {\xi}_R(t) {\xi}_R(t') \rangle = 2D_R \delta(t-t')$. For concreteness, the rotational diffusivity  $D_R$ is assumed to be related to the rotational friction coefficient $\gamma_R$ through the Smoluchowski-Einstein relation  $D_R = k_BT/\gamma_R$, with $T$ being the temperature and $k_B$ the Boltzmann constant. \rvsd{We also adopt the Stokes translational and rotational friction coefficients $\gamma_T = 6\pi \eta a_p$ and $\gamma_R = 8 \pi \eta a_p^3$   (note that the translational noise is of no significance at sufficiently high P\'eclet numbers that will be of interest here and is thus ignored in Eq. \eqref{vel_tr}; see, e.g., Refs. \cite{stark2019opt,Putzke2023}).}

\rvsd{Our modeling component (iii) enters through the term  $-\alpha\, \Omega(\mathbf{x}(t),\theta(t),t)$ in Eq. \eqref{vel_rot} and selecting the steering orientation in  Eq. \eqref{vel_tr} using the DRL approach. Before proceeding to discuss these aspects further (Section \ref{subsec:DRL}),} we note that, in the context of nonsmart ABPs, such a term can stem from a deterministic torque due, e.g., to an intrinsic particle chirality or an external stimuli or field. Examples include magnetic/gyrotactic torques experienced by dipolar active particles in external magnetic/gravitational fields. It may also correspond to the realignment rate due to klinotactic response in certain types of active particles; see Ref. \cite{shabanniya2020active} and the original reference cited therein for further details on the noted examples. 

\subsection{Equations of motion with DRL}
\label{subsec:DRL}

Within the DRL framework that we shall introduce later, a model sABP predicts its preferred orientation $ \mathbf{e}_{\mathrm{pref}}(t)=(\cos\theta_{\mathrm{pref}}(t), \sin\theta_{\mathrm{pref}}(t))$ at time $t$ by requesting an A2C network  to  steer it toward the destination point. Here,  $\theta_{\mathrm{pref}}(t)$ is the predicted preferred orientation angle relative to the $x$-axis. \rvsd{This angle is not necessarily the optimal one $\theta_{\text{opt}}(t)$ that would produce the shortest-time trajectory.} With further details provided in Sec. \ref{subsec:UtRL}, we proceed in this section by discussing how the DRL is incorporated into the equations of motion. 

By defining the angular fluctuations variable $\psi(t) = \theta(t) - \theta_{\mathrm{pref}}(t)$, we adopt a reorientation mechanism similar to Ref. \cite{stark2019opt} by setting $\Omega(\mathbf{x}(t),\theta(t),t) = \psi(t) $. We  arrive at the following Langevin equation for the time evolution of angular fluctuations:
\begin{equation}\label{langevin}
	\dot{\psi}(t) = -\alpha\psi(t) + \xi_R(t),  
\end{equation}
\rvsd{where the realignment strength is restricted as $\alpha>0$.} 

\rvsd{The numerical implementation of the DRL using Eq. \eqref{langevin} involves two conceptual stages: the `training stage' and the `simulation stage'. During the training stage, the sABP continually strives to find better policies. In our case, this means finding trajectories of shorter and shorter navigation times, until the optimal policy equivalent to predicting optimal steering angles is achieved, i.e., $\theta_{\text{pref}}(t) \simeq \theta_{\text{opt}}(t)$. The obtained optimal policy in the training stage is then used to generate agent trajectories in the simulation stage. As noted before, the simulated  trajectories terminate when the agent hits the target wall  (Fig. \ref{Gaus_rot}). This differs from the situation in the training stage, where agent trajectories are terminated when certain criteria, as specified in Sec. \ref{subsec:UtRL}, are satisfied.}

\rvsd{While Eq. \eqref{langevin} is similar to the one used in Ref. \cite{stark2019opt}, our approach is based on DRL  and differs from the RL approach used in the mentioned reference. The specific differences between our methods and those of Ref. \cite{stark2019opt}, can be summarized as follows. First, within the tabular $Q$-learning algorithm in Ref. \cite{stark2019opt}, the thermal noise is not applied during the training stage, where $\theta_{\text{opt}}(t)$ is obtained. It is applied only during the simulation stage. Our approach, however, allows for the noise to be included in both the training and the simulation stages of the implementation. As we shall see, the noise strength may also differ between the training stage and the simulation stage. Second, the algorithm used in Ref. \cite{stark2019opt} is best suited for discretized (lattice) backgrounds, where the plane of motion is divided into a finite grid of cells. By contrast, our approach relies on continuous prediction of the shortest-time trajectory across the plane of motion, providing a more flexible and accurate solution especially for dynamic environments.}

\subsection{Dimensionless representation}
\label{subsec:quant}

To avoid redundancies in the parameters and to simplify the equations of motion, we introduce dimensionless quantities by non-dimensionalizing the scales of length with a reference diameter $d$ for the sABPs; that is,  $x \rightarrow x/d$ and $y \rightarrow y/d$. \rvsd{The other parameters are rescaled as  $t\rightarrow tv_0/d$,  $a_p \rightarrow a_p /d$,  $L \rightarrow L /d$, $U\rightarrow U/U_0$, $|\mathbf{x}_c(t)| \rightarrow |\mathbf{x}_c(t)| /d$, $R_c \rightarrow R_c /d$, $\omega\rightarrow \omega d/v_0$, $\gamma_T\rightarrow \gamma_T v_0d/U_0$, $\gamma_R\rightarrow\gamma_R v_0 /(k_B T\,d) = v_0 /(D_R\,d)$. The latter gives the rescaled rotational diffusivity $D_R\rightarrow D_R d/v_0$, which is nothing but the inverse P\'eclet number $1/\text{Pe}$, with the definition} 
\begin{equation}
\label{eq:Pe}
\text{Pe} = v_0/(D_R d).
\end{equation}
Equation \eqref{vel_rot} can then be written in non-dimensional form and for components $x(t)$ and $y(t)$ as 
\begin{equation}\label{mot_dimless_x}
     \dot{x}(t) = \cos\theta(t) - \gamma_T^{-1}  \frac{\partial U(x,y,t)}{\partial x},
\end{equation}
\begin{equation}\label{mot_dimless_y}
     \dot{y}(t)  = \sin\theta(t) - \gamma_T^{-1}   \frac{\partial U(x,y,t)}{\partial y}. 
\end{equation}
Upon rescaling, Eq. \eqref{langevin} remains in the exact same  by redefining  $\alpha \rightarrow \alpha d/v_0$. Also, the noise term in Eq. \eqref{langevin} is rescaled as $\xi_R\rightarrow  \xi_R d/v_0$ and its correlator as $\langle\xi_R(t) \xi_R(t') \rangle =  2 \text{Pe}^{-1} \delta(t - t')$. \rvsd{The rescaled parameter $\alpha$ can be varied to study different realignment strengths but, for the sake of clarity, we set $\alpha=1$ (note the rescaled units) to reduce the number of free parameters in our analysis.}

\rvsd{In addition to  the rescaled particle radius $a_p$ and the P\'eclet number $\text{Pe}$, the set of dimensionless parameters for the present system includes the rescaled box size $L$, the rescaled peak location and range of the potential barrier, $|\mathbf{x}_c(t)|$ and $R_c$, respectively, as well as the ratio $6\pi \eta d^2 v_0/U_0$, which is embedded in the rescaled $\gamma_T$ and can be viewed as the rescaled viscosity of the background fluid.  In our forthcoming analysis, we shall vary $a_p$ and $\text{Pe}$, while keeping the remaining dimensionless parameters fixed (see Sec. \ref{subsec:parameters}). Specifically, we shall set $6\pi \eta d^2 v_0/U_0=1$. Our choices of rescaled parameters can be mapped to a wide range of realistic values for the system parameters in actual units. For instance, for sABP of diameter in the range \( a_p\in [0.1,1]\, \mu\text{m} \) operating in an aqueous medium of viscosity \( \eta = 0.001\, \mathrm{Pa \cdot s} \), with potential strengths of $U_0\in [10,100]\,k_BT$ (sufficiently larger than the thermal energy), the self-propulsion velocity is mapped to the experimentally feasible  \cite{bechinger2016ap} range of  \( v_0 \in [0.2,\, 200]\, \mu\text{m/s} \).}

\section{Methods}\label{sec:methods}

\begin{figure*}[t]
\centering
\includegraphics[width = .9\textwidth]{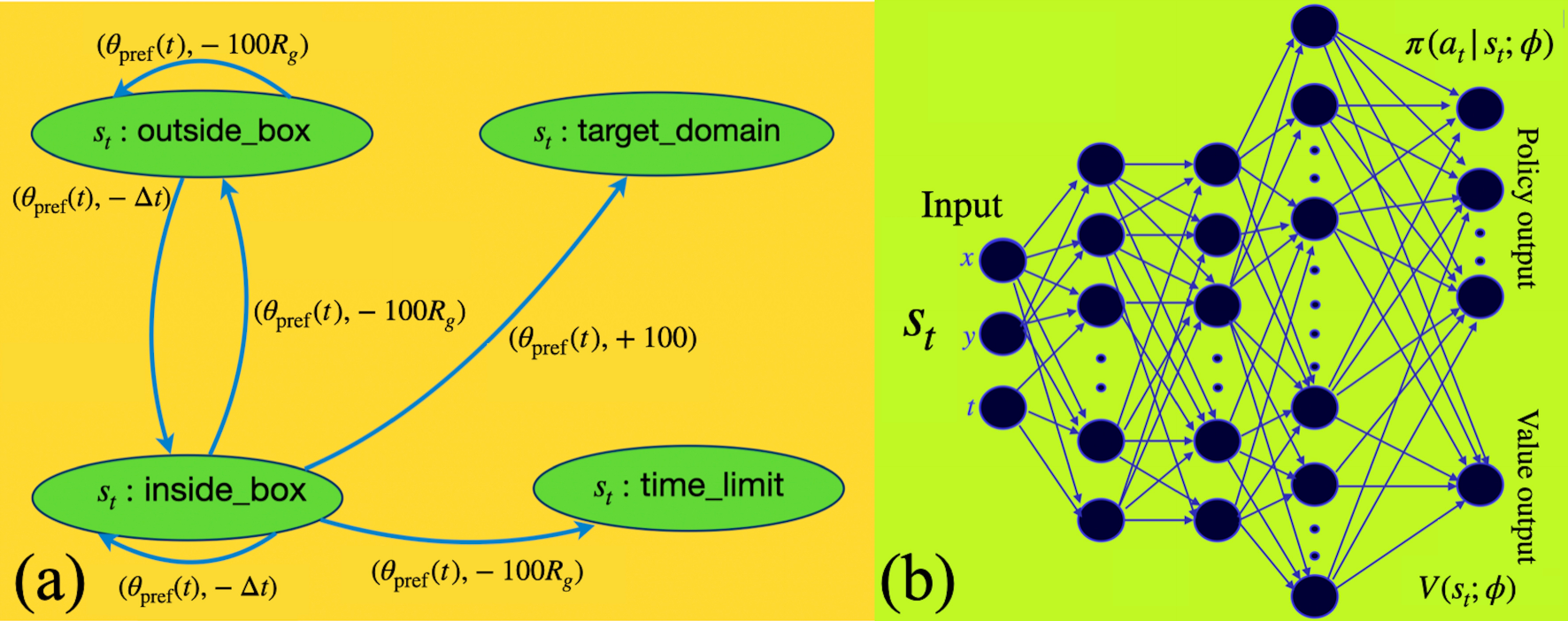}
\caption{(a)  MDP diagram for optimal navigation in the shortest time. Each edge demonstrates a pair of actions and rewards $(a_t,r_t)$ obtained by transitioning from state $s_t$ to a new state $s_{t+\Delta t}$. Also, four classes of states are represented by different labels: ``inside\_box'', ``outside\_box'', ``time\_limit'' and ``target\_domain''. (b) The A2C network diagram with input layer $s_t$, body and output layers as represented by the policy $\pi(a_t|s_t)$ and value $V(s_t)$ output functions. See the text for more details.
}
\label{MDP}
\end{figure*}

\subsection{Utilizing DRL for navigating sABPs}\label{subsec:UtRL}

In our implementation, the sABP interacts with its surrounding environment by perceiving (or, observing) the physical position as its {\em state}  at time $t$, $s_t = (x(t),y(t))$. Each state $s_t$ belongs to the space $\mathcal{S}$ of all possible positional states within the simulation box. When perceiving the state $s_t$ at time $t$, the sABP selects {\em action} $a_t$ by picking a steering angle $\theta_{\text{pref}}(t)$  from all possible angles $\mathcal{A} = \{m\pi/l | m = 0,1,2,...,l\}$ at angular separation $\pi/l$, where $l>0$ is a given angular resolution parameter that allows for fine-tuning between coarse and precise navigation controls. 

 To implement the aforesaid action, we update the direction of self-propulsion at the beginning of each {\em prediction} time step ($\Delta t$) as follows: $\mathbf{e}(t) \leftarrow \mathbf{e}_{\text{pref}}(t)$, where $\mathbf{e}_{\text{pref}}(t) = (\cos{\theta_{\mathrm{pref}}(t)},\sin{\theta_{\mathrm{pref}}(t)})$. This process involves integrating the translational and rotational equations of motion in Eqs. \eqref{vel_tr} and \ref{vel_rot} to determine the new state of the sABP in the next state $s_{t+\Delta t}$. \rvsd{The agent then receives a {\em reward} signal,  $r_t$, from the environment at the end of the prediction time step $\Delta t$. The mentioned numerical integration is carried out through {\em simulation} time steps ($\delta t$), much shorter than the prediction time step ($\delta t \ll \Delta t$). Throughout the integration process within the prediction time step, we track the exerted force $\mathbf{F}(t)$ until reaching the next state of the agent at $t+\Delta t$.} 

The preceding decision-making process (performed by selecting the desired actions) is modeled as a Markov Decision Process (MDP) \cite{sutton2018reinforcement}, a mathematical framework that provides a formalization for stochastic sequential decision-making problems. This is shown schematically in panel (a) of Fig. \ref{MDP} where the states are depicted as closed ellipses and the pairs of action and reward $(a_t,r_t)$ by interconnecting directed arrows. 

The provided MDP diagram has two classes of states: non-terminal and terminal. \rvsd{The non-terminal states labeled by ``inside\_box'' represent the states where the agent is inside the simulation box, and ``outside\_box'' indicates states where the agent crosses the boundaries and moves outside the simulation box.} Conversely, the terminal states are labeled ``time\_limit'' and ``target\_domain''. The ``time\_limit'' state represents the scenario where the agent fails to reach the target domain within the specified time limit $T_{\text{max}}$. The ``target\_domain'' state indicates that the agent has successfully reached the target domain at the destination point G (see Fig. \ref{Gaus_rot}) in less time than $T_{\text{max}}$. Through this MDP process, we have the following rewarding system: Movements are determined by selecting actions $a_t$ at each time step within the simulation box, resulting in a penalty for our agent of $r_t = -\Delta t$, \rvsd{where $\Delta t$ is the prediction time step required for the transition to the next consecutive state}. In case of being in states labeled by ``outside\_box'' and ``time\_limit'', we penalize the agent by $r_t = -100R_g$ where $R_g$ is defined as the in-plane distance between the terminal state outside the target domain and the destination point (this penalty adjustment are especially useful to avoid local optima and reach more rapidly to the shortest path). In the MDP diagram, transitioning to the target domain is awarded by a huge positive value of $+100$ that helps the sABP discover the target domain. 

By formulating our problem within an MDP and utilizing DRL as we shall discuss next, we predict the approximate optimal {\em policy} function that can guide the agent to reach the target domain in the shortest possible time. 

\subsection{Training sABP using A2C method}\label{subsec:train_a2c}

We implement the Advantage Actor-Critic (A2C) method as our DRL approach of choice to train the agent. This method merges the strengths of value-based and policy-based RL approaches. We closely follow the  A2C methodology and implementation of  Ref. \cite{Rev_motion_planning,gunnarson2021learn,nasiri2022reinforcement,morales2020grokking} together with the PyTorch library \cite{paszke2019pytorch}  \rvsd{(some of the basic notions of the A2C method are summarized in App. \ref{appendix:DRL_basics})}. The A2C method is structured around two key components, as demonstrated in panel (b) of Fig. \ref{MDP}: The actor network, denoted by $\pi(a_t | s_t; \phi)$, and the critic network, denoted by $V(s_t; \phi)$. Both the actor and critic networks are parameterized by a shared set of parameters $\phi$, creating a unified network to obtain the optimal policy and the value function. 

In the training stage, the agent  continually attempts to obtain the optimal policy in sequential episodes via the A2C method. Each episode in the context of RL starts at the initial state (which, in the present context, corresponds to the starting point $\mathrm{S}$; see Fig. \ref{Gaus_rot}) and ends when the termination criteria are fulfilled (see Sec. \ref{subsec:UtRL}). The first episode is initially started with the A2C network being randomly weighted. Through each episode of learning, based on the predictions obtained by the A2C network, we generate the fixed number of trajectories ($N$) following the policy $\pi(a_t|s_t;\phi)$, where the parameter set $\phi$, determines the current weights of the network. The agent generates trajectories $\tau_\phi$ that are time-sequential sets of states and actions ($s_0,a_0,s_1,a_1,...,s_T$), starting from the initial state $s_0$ and terminating in the final state $s_T$, aiming at maximizing the cumulative rewards. 

To obtain the optimal value and policy functions (which give us the maximum cumulative rewards), we compute the losses $L_\pi(\phi)$ and $L_V(\phi)$, as collectively averaged across these trajectories. These aggregated loss values are then utilized for back-propagation to update the weights of the interconnected A2C network. The \rvsd{main objective  of the training} is to minimize the total loss $L(\phi) = L_V(\phi) + L_\pi(\phi)$; \rvsd{see App. \ref{appendix:DRL_basics}}.

\begin{figure*}[t]
\includegraphics[width=0.70\textwidth]{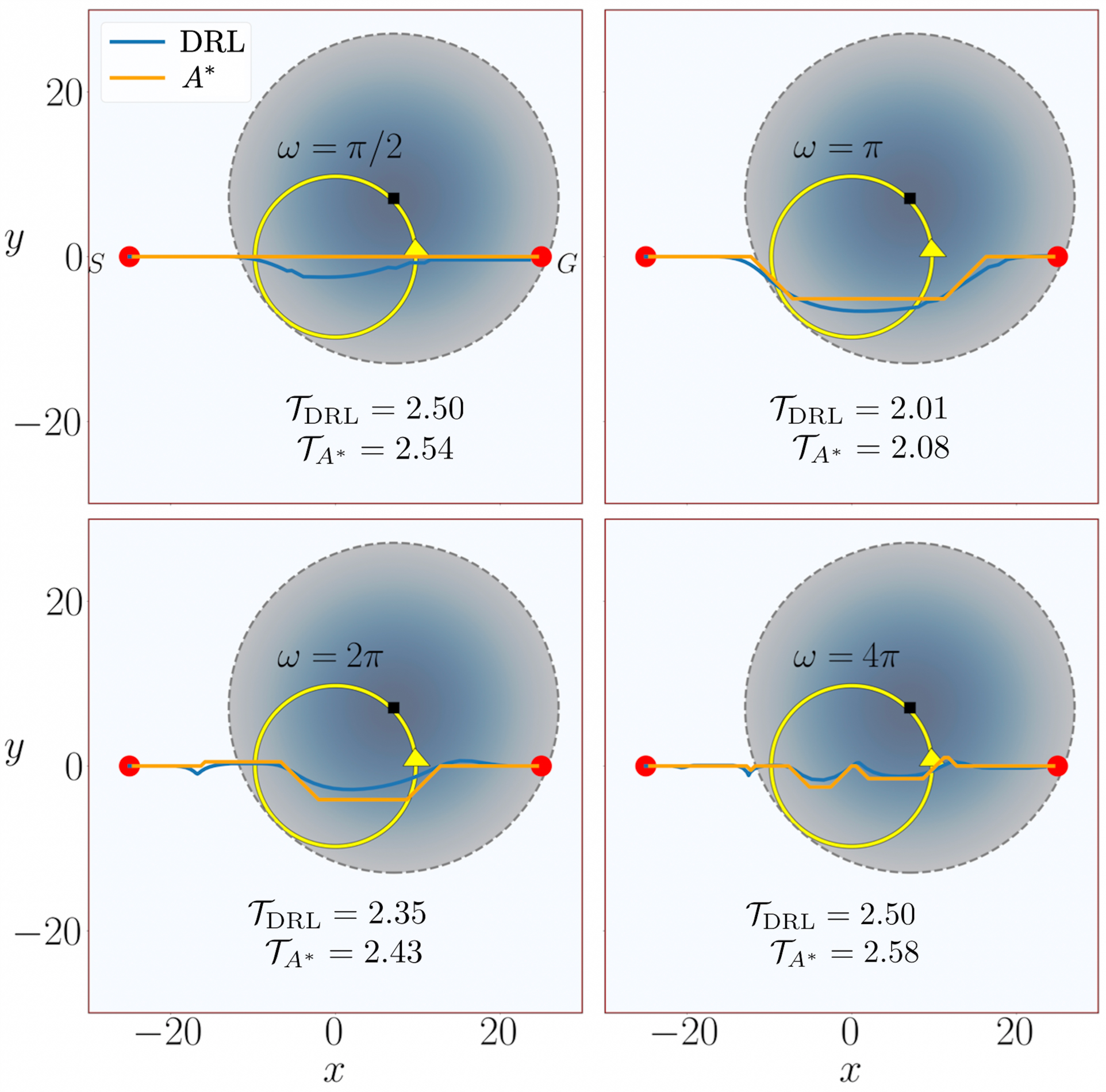}
\caption{\rvsd{Shortest-time trajectories of the modeled microagent (sAPB) moving from the start point, S, to the destination point, G (red bullets), across a repulsive Gaussian potential barrier. The center of the barrier rotates in a counterclockwise direction along the yellow circle with angular velocities \(\omega = \{\pi/2, \pi, 2\pi, 4\pi\}\), as indicated on the graphs.  The dark blue region shows the snapshot of the potential barrier  at times $t = \{1/2,1/4,1/8,1/16\}$, respectively. The blue sABP trajectories are obtained through noise-free DRL simulations with noise-free training. The orange trajectories show the shortest-time trajectory obtained by the \(A^*\) algorithm. The corresponding shortest arrival times  are represented in the insets of the graphs by $\mathcal{T}_{\text{DRL}}$ and $\mathcal{T}_{A^*}$.}
}
\label{pred_control}
\end{figure*}

\subsection{Choice of parameters for \rvsd{the training and simulation}}
\label{subsec:parameters}

\rvsd{As noted in Sec. \ref{subsec:quant}, we fix a number of the rescaled system parameters in our upcoming analysis. Specifically, we set the simulation box length as $L=$ 26, and assume that the sABP starts its navigation at  $(x, y) = (-25, 0)$. The target domain is taken as a circular region centered at $(x, y) = (25, 0)$ with a radius of $a_p/2$, which is \rvsd{half the radius} of the sABP. For the potential barrier, we set $|\mathbf{x}_c(t)| = R_c/2 = 10$, yielding a cutoff radius of $R_c = 20$. The time limit is set as  $T_\text{max} = 20$. These parameter values are used in both the training and simulation stages. However, in the training stage, we also fix the rescaled sABP radius as $a_p = 1$, sufficiently smaller than the box size, while we allow for $a_p$ to vary in the simulation stage.}

\rvsd{The P\'eclet numbers in the training and simulation stages, which we shall later denote by $\text{Pe}^{\text{sim}}$ and $\text{Pe}^{\text{train}}$, are varied as well. Hence, we shall explore the outcomes of the A2C deep reinforcement learning in achieving the shortest-time trajectories in the current setting across the parameter space, spanned by the particle size $a_p$ in the simulation stage as well as $\text{Pe}^{\text{train}}$ and $\text{Pe}^{\text{sim}}$. Recall that these two distinct P\'eclet numbers identify the thermal noise strengths of the background environment in the training and simulation stages, as implied by Eq. \eqref{eq:Pe}.}

We train the sABPs over 20,000 episodes in our implementation, averaging 10 trajectories per episode. \rvsd{We set the well-known hyperparameters of the A2C network \cite{morales2020grokking}, denoted by $\{\chi, \beta, \xi, \gamma\}$, as $ \chi = 10^{-4}, \beta = 10^{-3}, \xi = 0.5$ and $\gamma = 0.99$. We also set $\Delta t = 5 \times 10^{-2}$, $\delta t = \Delta t /50$  and $l = 64$. Hence, at each prediction time step $\Delta t$ (Sec. \ref{subsec:UtRL}), the A2C network determines the direction of $\mathbf{e}_{\text{pref}}(t)$ out of $l = 64$ possible steering angles.}

\rvsd{In what follows, the reported data are obtained by averaging the simulation results over $10^4$ independent trajectories produced using the A2C method.}

\begin{figure*}[t]
\centering
\includegraphics[width=.9\textwidth]{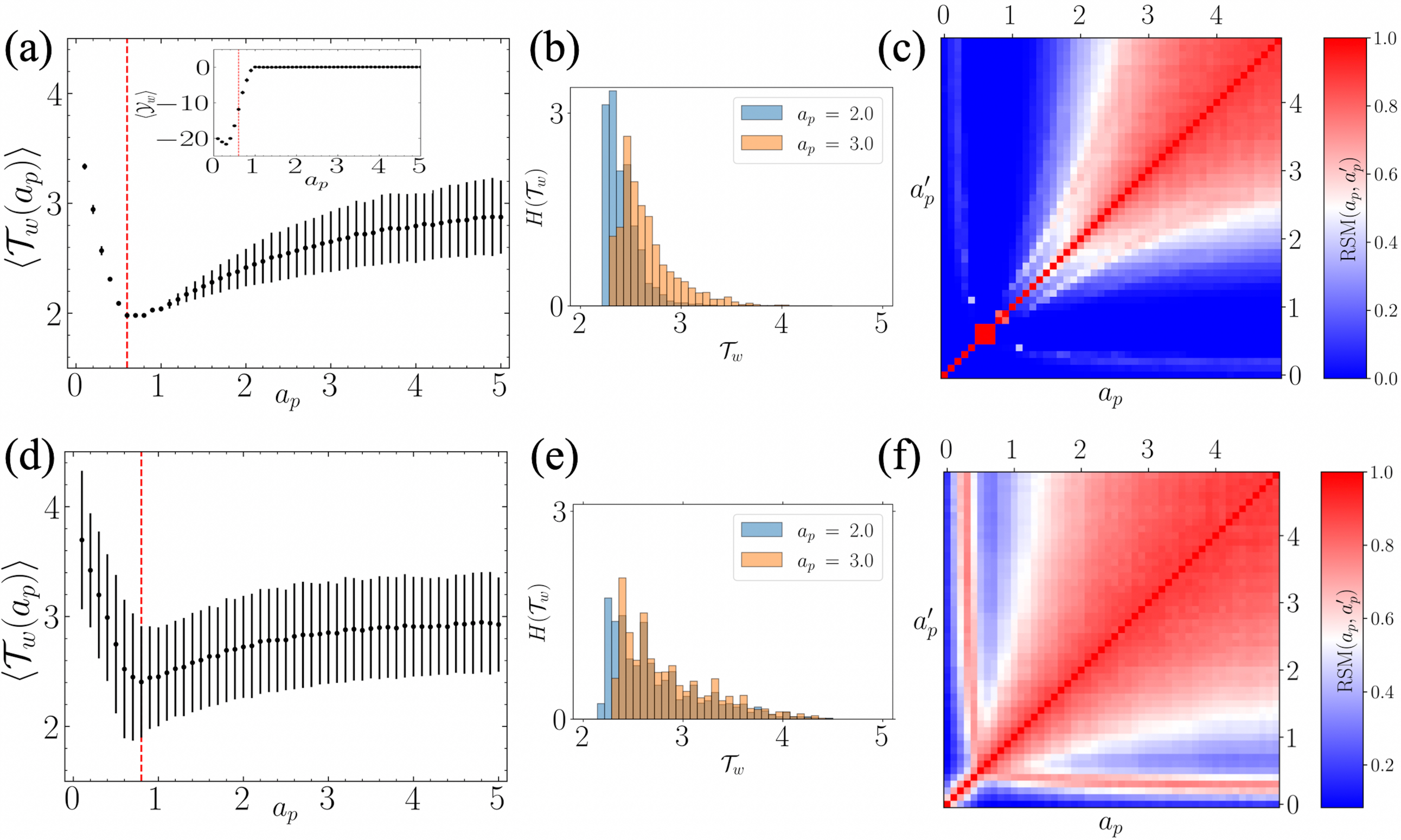}
\caption{\rvsd{(a) The mean wall-arrival time $\langle
		\mathcal{T}_{w}\rangle$ as a function of the sABP radius, $a_p$, in  noise-free simulations with noise-free training. The inset shows the mean lateral position (mean lateral diversion) of the sABPs, $
		\langle\mathcal{Y}_{w}\rangle$, upon arrival at the wall at $x=25$ as a function of $a_p$. The vertical, dashed, red line indicates the radius $a^*_p$ for which the wall-arrival time takes its minimum value. (b) The histogram of wall-arrival times $
		\mathcal{T}_{w}(a_p = 2.0, \text{blue})$ and $
		\mathcal{T}_{w}(a_p = 3.0, \text{orange})$. (c) The color map of $
		\mathrm{RSM}(a_p,a_p')$ as a function of $a_p$ and $a_p'$. Panels (d), (e), and (f) are the same as panels (a), (b), and (c), respectively, but they show the results for simulations in a noisy background with $	\text{Pe}^{\text{sim}} = 10^2$ and noise-free training. In all panels, we have  $\alpha = 1$ (as noted after  Eq. \eqref{mot_dimless_y}), $\omega = \pi$, and the other system parameters are according to Sec. \ref{subsec:parameters}.}}
\label{size-separation}
\end{figure*}

\section{Results and Discussions}
\label{sec:results}

\subsection{\rvsd{Shortest-time trajectories in the absence of noise}}
\label{subsec:trajectories}

\rvsd{In this section, we proceed by ignoring the thermal noise in the simulation and training stages within the A2C method. In other words, the P\'eclet number  $\text{Pe}^{\text{sim}}$ and $\text{Pe}^{\text{train}}$ are both infinitely large; see Eq. \eqref{eq:Pe}.}
\rvsd{In Fig. \ref{pred_control}, we show typical examples from the obtained shortest-time trajectories of microagents with $a_p = 1$ (shown as blue curves) at fixed barrier rotational velocities $\omega =\{\pi/2,\pi,2\pi,4\pi\}$ (see Secs. \ref{subsec:quant} and \ref{subsec:parameters} for parameter definitions and the parameter values that are fixed throughout our analysis).} To ensure the consistency of our results in obtaining optimal policies, we use multiple training sessions, each of them starting from an initial episode and terminating at a final episode. This will allow us to verify that the optimal policies are uniquely achieved across different training sessions. 

Additionally, we validate the optimality of our results  by comparing them with those we obtain from a graph-based algorithm $A^{*}$, shown as orange trajectories in Fig. \ref{pred_control}.   The \(A^{*}\) algorithm is implemented following the procedure described in Ref. \cite{Kularatne:2018aa} for finding the shortest-time trajectory in a dynamical background, which we have adapted for the case of sABPs. \rvsd{The close agreement found between A2C and $A^{*}$ results (Fig. \ref{pred_control}) serves as an important benchmark, ensuring that the A2C method indeed converges to the globally optimal shortest-time trajectories, rather than being trapped in suboptimal local minima in the absence of noise.}

\subsection{Size-based particle sorting \rvsd{in the absence of noise}}
\label{subsec:sizedet}

\rvsd{We proceed by examining the size-dependent behavior of the agent's arrival time $\mathcal{T}_{w}$ at the target wall,  extending vertically across the target domain along the $y$-axis at $x=25$ (vertical dashed line in Fig. \ref{Gaus_rot}). We fix the barrier rotational velocity as $\omega= \pi$. In the simulation stage, we discard the noise ($\text{Pe}^{\text{sim}}= \infty$), and vary the sAPB radius within  the discrete range $a_p =0.1, 0.2, 0.3,\cdots, 5$. We obtain $\mathbf{e}_{\mathrm{pref}}(t)$ from the A2C network trained for optimal navigation in the absence of noise and with the sABP radius chosen as $a_p = 1$, as mentioned before. Note that, in the simulations, the microagents are not necessarily required to arrive at the target domain, as their trajectories terminate once they hit the wall. This results in a lateral distribution of the termination points along the wall.} 

We plot the resulting mean arrival time at the wall,  $\langle\mathcal{T}_{w}\rangle$, as a function of the sAPB radius,  $a_p$, in panel (a) of Fig. \ref{size-separation}. \rvsd{The figure shows a nonmonotonic dependence for $\langle\mathcal{T}_{w}\rangle$ as a function of $a_p$. In other words, the mean wall-arrival time exhibits a  descending dependence on $a_p$ for $a_p < a^*_p$ and an ascending one for $a_p > a^*_p$, where $a^*_p$ (see the vertical, dashed red line) indicates the agent radius with minimum arrival time. Hence, two individual sABPs of different radii can be separated when their radii both fall within one of the mentioned size domains. The size-based  separation and sorting of the sABPs in this case are enabled by the distinct  arrival times of the sABPs at the wall, in each of the size domains $a_p < a^*_p$ and $a_p > a^*_p$.} 

\rvsd{If we consider a pair of  sABPs with radii $a_p' \neq a_p$ in the two separate size domains, $a_p' < a^*_p$ and $a_p > a^*_p$, one can still separate them based on the mean  lateral position  $\langle\mathcal{Y}_{w}\rangle$ where they hit the wall in $y$-direction. We will refer to    $\langle\mathcal{Y}_{w}\rangle$ as the {\em mean lateral diversion} of the sAPBs at the wall. This follows from the inset of Fig. \ref{size-separation}a where we have plotted  $\langle\mathcal{Y}_{w}\rangle$ as a function $a_p$, where $a^*_p$ is again indicated by a vertical, dashed red line. It is noteworthy that, for $a_p \ge 1$, the sABPs on average reach the target domain since $\langle\mathcal{Y}_{w}\rangle \simeq 0$ (Fig. \ref{size-separation}a, inset). In summary, a sample of sABPs with varying sizes can be separated by a rotating barrier and an embedded A2C network by utilizing their mean wall-arrival time $\langle\mathcal{T}_w\rangle$ and their mean lateral diversion at the wall, $\langle\mathcal{Y}_w\rangle$.}

\rvsd{However, as seen in Fig. \ref{size-separation}a, the error bars associated with the wall-arrival time grow as $a_p$ increases, indicating a larger degree of uncertainty in sorting the larger sABPs. This uncertainty is illustrated Fig. \ref{size-separation}b where histograms of the wall-arrival time for two sets of simulations with different particle radii, $a_p = 2$ and $3$, are shown (blue and orange, respectively). As the histograms overlap increases for different particle radii, so does the likelihood of misclassification and error in sorting particles at the target wall.}

\rvsd{To quantify the potential sorting error and the effectiveness of the size-sorting mechanism, we analyze the co-occurrence of wall-arrival times for sABPs of different sizes by examining how often their wall-arrival times fall into the same histogram bins. This is done by defining a {\em relative separability measure}  (RSM)  as}  
\begin{equation}
	\label{rsm_eq}
\rvsd{	\mathrm{RSM}(a_p,a'_p) = \sum_{i=1}^{n} \min\Big\{ H_i\big(\mathcal{T}_w(a_p)\big),\; H_i\big(\mathcal{T}_w(a'_p)\big) \Big\},}
\end{equation}
\rvsd{where  \(H_i\big(\mathcal{T}_w(a_p)\big)\) and \(H_i\big(\mathcal{T}_w(a'_p)\big)\) represent the wall-arrival histogram counts for particles of  sizes \(a_p\) and \(a'_p\) that arrive at the wall within the \(i\)-th time interval. Here, \(n\) is the total number of histogram bins representing different time intervals. We set \(n = 201\) and define the width of the bins as \(({T_{\text{max}} - T_{\text{min}}})/n\), where \(T_{\text{max}}=20\), as noted before, and  \(T_{\text{min}}\) is the minimum wall arrival time obtained from simulations for a given $a_p$. However, to have equal bin widths for different $a_p$, we take \( T_{\text{min}} = \langle\mathcal{T}_{w} (a_p = 0.6)\rangle = 1.98\).  Also, the symbol `min' in Eq. \eqref{rsm_eq} indicates that the RSM is obtained by summing the minimum of the two histogram counts over all bins.}

\rvsd{Note that the higher the RSM value the greater is the overlap between the wall-arrival histograms of the two particles and, hence, the lower is the separation efficiency of the current method for the given particle sizes. Figure \ref{size-separation}c shows a 2D color map of the RSM function in terms of  \(a_p\) and \(a'_p\). As the RSM function approaches zero, the color coding in the graph tends to blue, and as it approaches one, it tends to red. The red (\rvsd{blue}) color generally indicates a less (more) efficient separability of particle sizes  \(a_p\) and \(a'_p\).}

\rvsd{It is important to emphasize that the sorting mechanism described above is directly enabled by the choice of a rotating barrier. If we assume a static barrier ($\omega = 0$), the mean wall-arrival time shows a broad plateau across different values of $a_p$, making size-based separation impossible; see App. \ref{appendix:irrvsrot} for further details.}

\subsection{Size-based sorting \rvsd{in noisy background with noise-free training}}
\label{subsec:sizedetthermal}

\rvsd{In the preceding sections, we ignored the thermal noise during the simulation and training stages within the A2C method. We now include the noise term in the simulation stage. This is represented by a finite simulation P\'eclet number $\text{Pe}^{\text{sim}}$; see Eq. \eqref{eq:Pe}. Different values of  $\text{Pe}^{\text{sim}}$ represent different noise strengths. The agent's {\em training} is still performed in a {\em noise-free} background where the sABP radius is set to $a_p = 1$.}

\begin{figure*}[t!]
        \centering
        \includegraphics[width =0.825\textwidth]{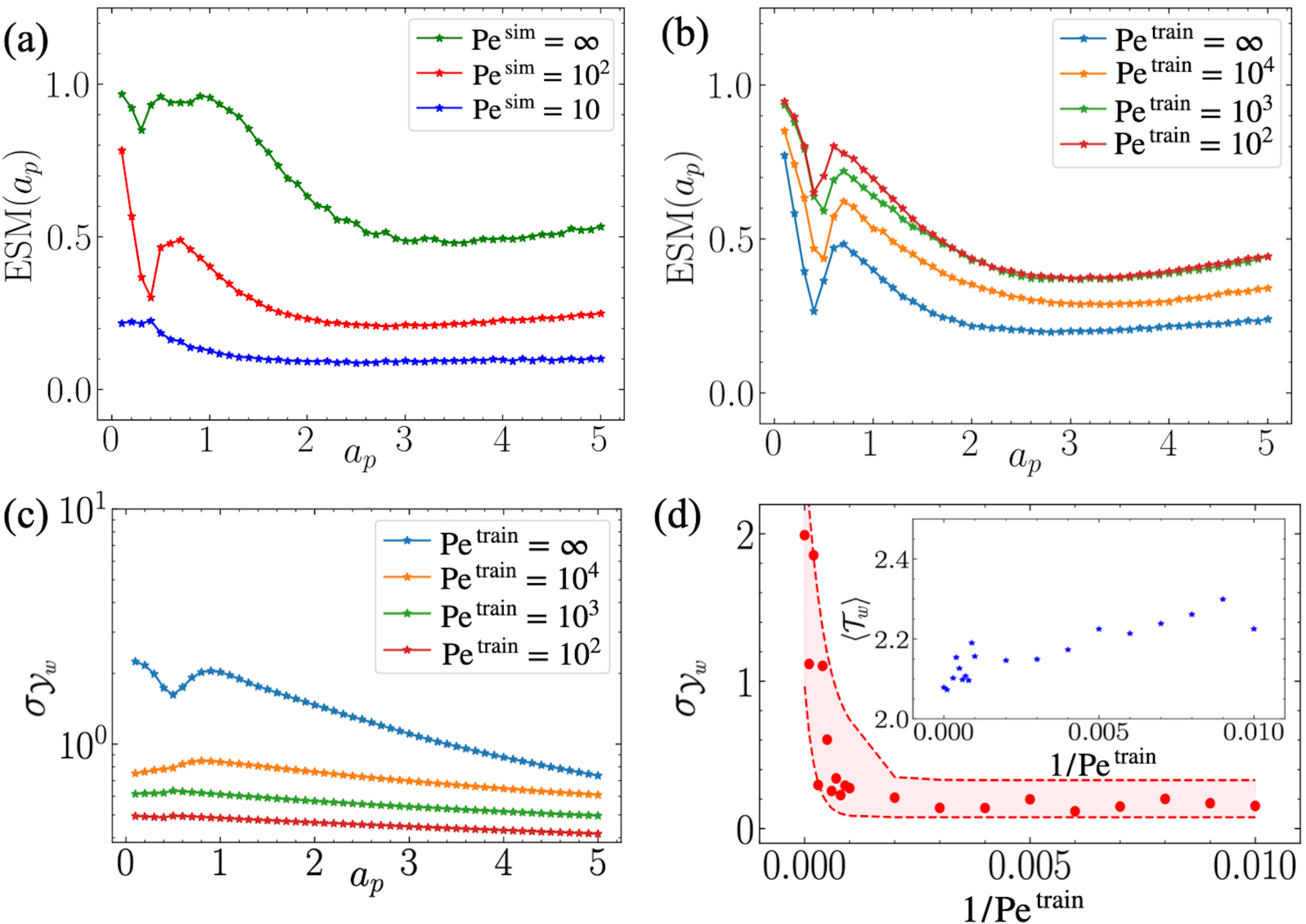}
        \caption{\rvsd{Panel (a)  shows the ESM obtained from simulations in a noisy background with noise-free training (Sec. \ref{subsec:sizedetthermal}), where the ESM values are evaluated across the size range $a_p=0.1,0.2,\ldots,5.0$ at fixed simulation P\'eclet numbers, $\text{Pe}^\text{sim}=\{\infty,10^2,10\}$, as indicated on the graph. Panel (b) is the same as (a) but here $\text{Pe}^\text{sim}=10^2$ is fixed and the ESM values are obtained by training the agents in a noisy environment (Sec. \ref{subsec:noisetrain}) using training Péclet numbers $\text{Pe}^\text{train}=\{\infty,10^{4},10^{3},10^2\}$. Panel (c) shows the standard deviation of the lateral diversion for sABPs corresponding to the curves shown in panel (b). Panel (d), main set, shows the standard deviation for unit agent  with $a_p=1$ as a function of $1/\text{Pe}^\text{train}$ at fixed $\text{Pe}^\text{sim}=10^2$. The inset shows the corresponding mean wall-arrival time, $\langle\mathcal{T}_w\rangle$, is  as a function of $1/\text{Pe}^\text{train}$. In all panels shown here,  $\alpha = 1$, $\omega = \pi$, and the other system parameters are the same as those in Sec. \ref{subsec:parameters}.}} 
        \label{noise-induced-test}
\end{figure*}

\rvsd{In Fig. \ref{size-separation}, panels (d), (e), and (f), we show the simulation results for  $\text{Pe}^{\text{sim}} = 10^2$. The particle radius $a_p$ is again varied within the range  $a_p= 0.1, 0.2,0.3,\cdots, 5$. Panel (d) clearly shows that the error bars in the mean arrival time at the wall  are enhanced across the whole size range investigated. The histograms for the wall-arrival times in panel (e) exhibit broadening from the right (larger $\mathcal{T}_w$ values) relative to the noise-free case in panel  (b). Accordingly, the RSM in a noisy background in panel (f) shows that the \rvsd{blue} areas shrink significantly relative to the noise-free case in panel (c). Thus, the efficiency of  size-based  sorting  in a noise background is largely diminished for agents trained in the absence of noise.} 

\rvsd{To shed further light on the noise-induced behavior of the sABPs, we introduce the {\em effective separability measure} (ESM) as follows
\begin{equation}
\mathrm{ESM}(a_p) = \frac{1}{N - 1}\sum_{a'_p \neq a_p} \big[1 - \mathrm{RSM}(a_p, a'_p)\big],
\end{equation}
where $N$ is the number of distinct particle sizes for which the size-based sorting is  considered, and the sum runs over \(N - 1\)  particle sizes  \(a'_p\) that are different from the selected size \(a_p\). The ESM for a specific size \(a_p\) is thus calculated by summing \((1 - \mathrm{RSM}(a_p, a'_p))\), which can itself be viewed as a measure of `successful' separation of size \(a_p\) from all other sizes \(a'_p\neq a_p\). Larger (smaller) ESM values for a given particle size indicate that the current mechanism is more (less) effective at separating that size from the others.}

In panel (a) of Fig. \ref{noise-induced-test}, we show the ESM values that are obtained from the \rvsd{2D maps of RSM computed within the simulations}. Different curves correspond to different values of $\text{Pe}^\text{sim} = \{\infty,10^2,10\}$ as indicated on the graph. Examining panels (a) and (d) of Fig. \ref{noise-induced-test}, we find that by decreasing $\text{Pe}^\text{sim}$, the ESM value shifts significantly towards lower values for all agent sizes analyzed. \rvsd{Therefore, as the noise strength gradually increases, it becomes more difficult to separate and sort agents of different sizes in the simulations using the current method.} 

\subsection{Size-based sorting \rvsd{for agents trained in noisy background}}
\label{subsec:noisetrain}

\rvsd{We now turn to the case where agents are trained in a noisy background. In other words, the P\'eclet number has a finite value not only in the simulation stage  (as in Sec. \ref{subsec:sizedetthermal}) but also in the training stage. We train the sABPs by restricting the P\'eclet number in the training stage to be equal to or larger than the P\'eclet number in the simulation stage; i.e., \( \text{Pe}^{\text{train}} \geq \text{Pe}^{\text{sim}} \). In other words, the training is done  in a {\em less} or {\em equally noisy} environment as the simulation.}  

\rvsd{Panel (b) of Fig. \ref{noise-induced-test} show how training in a noisy background affects the simulation results. In the plot, the P\'eclet number in the simulation stage is fixed as \( \text{Pe}^{\text{sim}} = 10^2 \) and the P\'eclet number in the training stage varies as \( \text{Pe}^{\text{train}} = \{\infty, 10^4, 10^3, 10^2\} \). Note that the blue curve in panel (b) is the same as the red curve in panel (a), representing the case with noise-free training ($\text{Pe}^{\text{train}} = \infty$).} 

\rvsd{It is evident from panel (b) that, as the  \( \text{Pe}^{\text{train}} \) decreases toward  \( \text{Pe}^{\text{sim}} \), the ESM increases across all the investigated values of $a_p$. Hence, not only does the efficiency of size-based separation improves relative to the noise-free case but the efficiency generally improves when the training background becomes noisier (corresponding to lower $\text{Pe}^{\text{train}}$).}

\rvsd{Our results in panel (c) of Fig. \ref{noise-induced-test} shows that decreasing the training P\'eclet number also leads to a significant reduction in the standard deviation  \( \sigma_{\mathcal{Y}_{w}} \)  of the lateral diversion of sABPs, as they reach the wall. The mentioned standard deviation  is calculated from termination latitudes of individual agent trajectories at the target wall. In other words, the termination latitudes exhibit a smaller spread when the agents are trained in a background with comparable strength of noise as in the simulations. Panel (d) of Fig. \ref{noise-induced-test} shows the more detailed behavior of \( \sigma_{\mathcal{Y}{w}} \)  as a  function of  \( 1/\text{Pe}^{\text{train}} \)  for the case with $a_p = 1$. Here, simulation data are shown by solid red circles and the dashed curves merely serve as guides to eye, bracketing the data in between. As seen,  \( \sigma_{\mathcal{Y}{w}} \)  nearly converges to a plateau and becomes independent of  the training P\'eclet number   for   \( 1/\text{Pe}^{\text{train}}>0.002 \), or \( \text{Pe}^{\text{train}}<500 \). On the other hand, inspecting the mean wall-arrival times \(\langle \mathcal{T}_w \rangle\) in the inset of panel (d) indicates that it  also takes longer for the sABPs to reach the wall when the noise intensity used in the training stage approaches  the noise intensity of the simulation stage.} 

\section{Concluding remarks}
\label{sec:conclusions}

\rvsd{In this study, we have investigated the application of  DRL using a A2C method for the optimal navigation of smart active Brownian particles (microagents) through a dynamic environment. The latter features a repulsive Gaussian potential that rotates around a fixed center. We have also examined the optimal microagent navigation in noise-free and noisy backgrounds during the training and simulation stages. We have validated the approach through extensive simulations, confirming the robustness and efficiency of the proposed navigation strategies. Additionally, we have explored the implications of our method for size-based separation of microagents. By using the mean wall-arrival times besides the mean  lateral diversion of the microagents at the target wall,  and additionally introducing a novel relative separability measure, we have demonstrated that the microagents can be separated and sorted based on their sizes.}

\rvsd{When the navigation occurs in the presence of noise, training the microagents in a noisy background reveals a significant improvement in their size-based sorting compared to the cases where the training is done in a noise-free environment. This is established by defining an effective separability measure to quantify the sorting efficiency in noisy environments. These findings highlight the potential of noise-assisted training to enhance the robustness of DRL-based navigation strategies under realistic conditions.}

\rvsd{As demonstrated through our results, the rotating nature of the potential barrier plays a key role in enabling size-based separation of microagents, given that the separation is entirely precluded when the barrier is static (App. \ref{appendix:irrvsrot}).}

\rvsd{It should be noted that our model is based on certain simplifying assumptions. In particular,  advective flows and hydrodynamic effects that may arise from the fluid disturbance due to the agent or barrier motion have not been taken into account. Also, our adoption of Stokes friction coefficients and the Smoluchowski-Einstein between rotational diffusivity and friction coefficient introduce additional simplifications in describing the stochastic motion of smart microagents. In the case of anisotropic particles, particle shape factors can lead to rather cumbersome forms for the aforementioned quantities than what is assumed in our model \cite{kim2005microhydrodynamics,happel1983low}. Also,  the effective rotational diffusivity in the case of active particles may be fundamentally different from the rotational Stokes diffusivity because of intrinsic, athermal, random processes that can  control the reorientation dynamics of these particles; see, e.g., Refs.  \cite{romanczuk2012active,romanovsky2013models}. In such cases, the translational and rotational diffusivities (or, the corresponding friction coefficients) can be unrelated; see, e.g., Ref. \cite{li2020diffusion}. Finally, the specific form of the barrier used here (as a repulsive Gaussian potential) should be viewed as a toy model to illustrate the adaptability of DRL techniques to dynamically evolving environments. In experiments, we expect that such a potential barrier could be realized using, e.g., localized light-intensity fields produced by laser or even external magnetic fields applied locally to a sample containing magnetically responsive microagents. The barrier rotation around a fixed center can then be generated by moving the external source. In general, the navigation strategies proposed here are expected to remain applicable to real-life microscale systems (such as microfluidic setups), provided that the models used for the microagent and environment are properly refined to capture the relevant details of the specific systems in question.}

\begin{figure*}[t!]
        \centering
        \includegraphics[width=.9\textwidth]{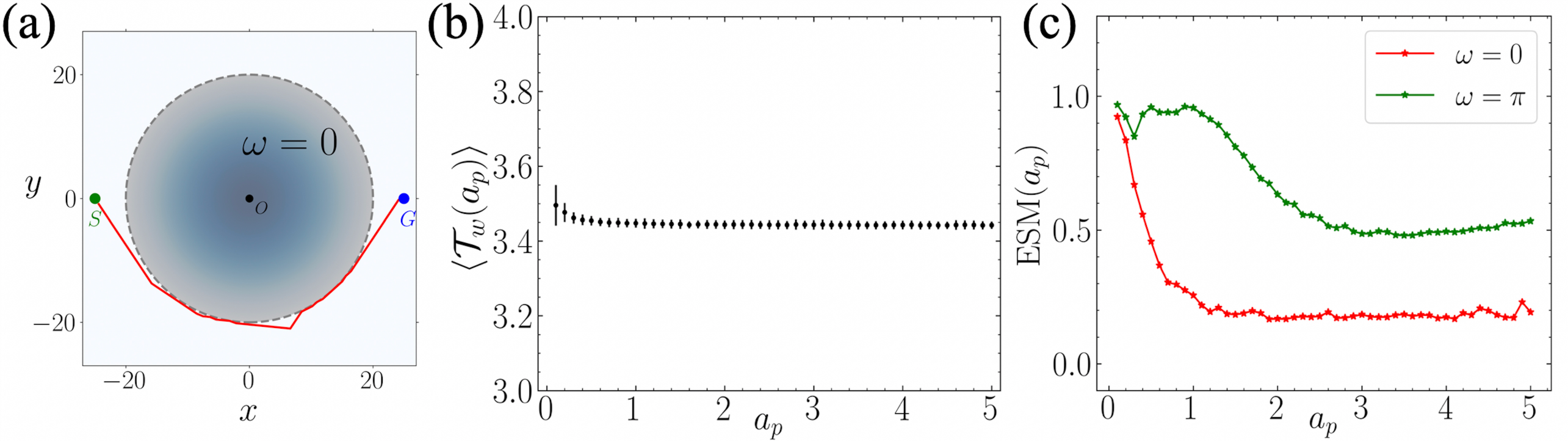} 
        \caption{\rvsd{(a) The red curve shows the shortest-time sABP trajectory obtained using the A2C method when the potential barrier is static (non-rotating, $\omega = 0$). Here, we have used   noise-free conditions in both training and simulation stages. (b) The corresponding mean wall-arrival time as a function of sABP radius. (c) The ESM values for the static barrier ($\omega = 0$) are compared with those in the case of a rotating barrier ($\omega = \pi$). The choice of system parameters are consistent with those reported in Sec. \ref{subsec:sizedet}.}}
        \label{fig:sorting}
\end{figure*}

\section*{Conflicts of interest}	
There are no conflicts of interest to declare.

\section*{Author contributions}

{\bf Mohammad Hossein Masoudi:} Conceptualization (supporting); Data curation (lead); formal analysis (lead); investigation (lead); methodology (lead); software (lead); validation (lead); visualization (lead); writing -- original draft preparation (lead); writing -- review and editing (equal). {\bf Ali Naji:} Conceptualization (lead); formal analysis (supporting); project administration (lead); funding acquisition (lead); investigation (supporting); supervision (lead); validation (supporting); writing -- original draft preparation (supporting); writing -- review and editing (equal). 

\section*{Acknowledgment}
A.N. would like to acknowledge support from the ICTP (Trieste, Italy) through the Associates Programme and from the Simons Foundation through grant number 284558FY19. MHM acknowledges the support provided through the same grant for a visit to the ICTP during this research.

\section*{Data availability statement}

The data supporting the results of this study are available within the article.	


%

\appendix

\section{\rvsd{Preliminaries of RL and A2C methods}\label{appendix:DRL_basics}}

\rvsd{Reinforcement learning is a framework for training an agent to make decisions in an environment by taking actions to maximize cumulative rewards (Ref. \cite{sutton2018reinforcement}). At each time step, the agent observes a state $s_t$, selects an action $a_t$, and receives a reward $r_t$, transitioning to the next state $s_{t+1}$. The cumulative reward, calculated till $t' = T_{\text{max}}$ and is defined as:
\[
G_t = \sum_{t'=t}^{T_{\text{max}}}\gamma^{t'-t}r_{t'},
\]
where $\gamma \in [0, 1]$ is the discount factor, controlling the emphasis on future rewards (see Ref. \cite{sutton2018reinforcement}).}

\rvsd{The RL methods can be categorized as  {\em value-based} and {\em policy-based methods}. The value-based methods focus on learning the value of states or state-action pairs. For instance, $Q$-learning estimates the action-value function $Q(a_t, s_t)$ and selects actions that maximize $Q$. The policy-based methods  directly optimize the policy $\pi(a_t | s_t)$ by maximizing the expected cumulative rewards.}

\rvsd{The A2C method is, on the other hand,  a deep reinforcement learning approach designed to employ both value-based and policy-based methods concurrently (see Ref. \cite{morales2020grokking} for further details). The essence of the A2C model lies in its advantage function, $A(a_t, s_t; \phi) = Q(a_t, s_t) - V(s_t; \phi)$, which evaluates the relative benefit of taking action $a_t$ in state $s_t$ denoted by the action-value function $Q(a_t, s_t)$ under the policy $\pi(a_t | s_t; \phi)$  compared to the baseline value provided by the critic network as the value function $V(s_t; \phi)$.  Here, the action-value function can be approximated through $Q(a_t, s_t) \simeq G_t$.}

\rvsd{The critic-network aims to minimize the squared advantage, formulated as $L_V(\phi) = \chi A^2(a_t, s_t; \phi)$, where $\chi$ is a coefficient that balances the emphasis on the value function loss. In parallel, the actor-network strives for policy optimization by minimizing its loss, $L_\pi(\phi) = -\beta \ln\pi(a_t | s_t; \phi) \cdot A(a_t, s_t; \phi) - \xi H(\pi(a_t | s_t; \phi))$, and the $\beta$ coefficient weigh the effect of policy losses. We also employ the entropy term, $H(\phi) = -\sum_{a_t} \pi(a_t | s_t; \phi) \ln \pi(a_t | s_t; \phi)$ weighted by $\xi$,  to encourage exploration and prevent the premature convergence to suboptimal policies. Both networks concurrently adjust their shared parameters $\phi$ through gradient descent: $\phi \leftarrow \phi + \nabla_\phi L_V(\phi)$ and $\phi \leftarrow \phi + \nabla_\phi L_\pi(\phi)$.}

\rvsd{The architecture of our A2C network is structured with three principal layers, following Ref. \cite{morales2020grokking}: an input layer that processes the spatial and temporal variables $x$, $y$, and $t$; a processing layer composed of two sublayers, and an output layer with two separate policy and value function outputs. The policy output is tasked with generating stochastic policies, $\pi(a_t | s_t; \phi)$, which calculates the probability of selecting each action $a_t$ from the set of possible actions. This involves choosing the steering angle from a set of discrete angles at the beginning of each time step. Simultaneously, the value output, $V(s_t; \phi)$, estimates the expected returns from state $s_t$, essential in computing the advantage function and guiding the policy towards more rewarding actions.}

\section{\rvsd{Static (non-rotating) potential barrier}}
\label{appendix:irrvsrot}

\rvsd{Panel (a) in Fig. \ref{fig:sorting} shows  the shortest-time trajectory of the microagent (red curve) when the  potential  barrier is static ($\omega=0$). Here, we use the noise-free training and simulation scenarios with the same choice of parameters as in Sec. \ref{subsec:sizedet}. As seen, the shortest-time trajectory follows the brim of the potential, qualitatively consistent with the results obtained in the case of a static Mexican-hat potential using $Q$-learning methods in Ref. \cite{stark2019opt} (see Fig. 4b therein). The mean wall-arrival time in panel (b) exhibits a  plateau through nearly the whole range of sizes considered here. It is thus evident that the sorting mechanism described in this work will be absent when the barrier is static. This is corroborated further with the highly suppressed ESM values shown in panel (c), where the results for the static case are compared with those for a rotating barrier with $\omega=\pi$.}

\bibliography{references}

\begin{thebibliography}{58}%
\makeatletter
\providecommand \@ifxundefined [1]{%
 \@ifx{#1\undefined}
}%
\providecommand \@ifnum [1]{%
 \ifnum #1\expandafter \@firstoftwo
 \else \expandafter \@secondoftwo
 \fi
}%
\providecommand \@ifx [1]{%
 \ifx #1\expandafter \@firstoftwo
 \else \expandafter \@secondoftwo
 \fi
}%
\providecommand \natexlab [1]{#1}%
\providecommand \enquote  [1]{``#1''}%
\providecommand \bibnamefont  [1]{#1}%
\providecommand \bibfnamefont [1]{#1}%
\providecommand \citenamefont [1]{#1}%
\providecommand \href@noop [0]{\@secondoftwo}%
\providecommand \href [0]{\begingroup \@sanitize@url \@href}%
\providecommand \@href[1]{\@@startlink{#1}\@@href}%
\providecommand \@@href[1]{\endgroup#1\@@endlink}%
\providecommand \@sanitize@url [0]{\catcode `\\12\catcode `\$12\catcode
  `\&12\catcode `\#12\catcode `\^12\catcode `\_12\catcode `\%12\relax}%
\providecommand \@@startlink[1]{}%
\providecommand \@@endlink[0]{}%
\providecommand \url  [0]{\begingroup\@sanitize@url \@url }%
\providecommand \@url [1]{\endgroup\@href {#1}{\urlprefix }}%
\providecommand \urlprefix  [0]{URL }%
\providecommand \Eprint [0]{\href }%
\providecommand \doibase [0]{https://doi.org/}%
\providecommand \selectlanguage [0]{\@gobble}%
\providecommand \bibinfo  [0]{\@secondoftwo}%
\providecommand \bibfield  [0]{\@secondoftwo}%
\providecommand \translation [1]{[#1]}%
\providecommand \BibitemOpen [0]{}%
\providecommand \bibitemStop [0]{}%
\providecommand \bibitemNoStop [0]{.\EOS\space}%
\providecommand \EOS [0]{\spacefactor3000\relax}%
\providecommand \BibitemShut  [1]{\csname bibitem#1\endcsname}%
\let\auto@bib@innerbib\@empty
\bibitem [{\citenamefont {Marchetti}\ \emph {et~al.}(2013)\citenamefont
  {Marchetti}, \citenamefont {Joanny}, \citenamefont {Ramaswamy}, \citenamefont
  {Liverpool}, \citenamefont {Prost}, \citenamefont {Rao},\ and\ \citenamefont
  {Simha}}]{marchetti2013hydrodynamics}%
  \BibitemOpen
  \bibfield  {author} {\bibinfo {author} {\bibfnamefont {M.~C.}\ \bibnamefont
  {Marchetti}}, \bibinfo {author} {\bibfnamefont {J.~F.}\ \bibnamefont
  {Joanny}}, \bibinfo {author} {\bibfnamefont {S.}~\bibnamefont {Ramaswamy}},
  \bibinfo {author} {\bibfnamefont {T.~B.}\ \bibnamefont {Liverpool}}, \bibinfo
  {author} {\bibfnamefont {J.}~\bibnamefont {Prost}}, \bibinfo {author}
  {\bibfnamefont {M.}~\bibnamefont {Rao}},\ and\ \bibinfo {author}
  {\bibfnamefont {R.~A.}\ \bibnamefont {Simha}},\ }\bibfield  {title} {\enquote
  {\bibinfo {title} {Hydrodynamics of soft active matter},}\ }\href
  {https://doi.org/10.1103/RevModPhys.85.1143} {\bibfield  {journal} {\bibinfo
  {journal} {Rev. Mod. Phys.}\ }\textbf {\bibinfo {volume} {85}},\ \bibinfo
  {pages} {1143--1189} (\bibinfo {year} {2013})}\BibitemShut {NoStop}%
\bibitem [{\citenamefont {Elgeti}, \citenamefont {Winkler},\ and\ \citenamefont
  {Gompper}(2015)}]{elgeti2015physics}%
  \BibitemOpen
  \bibfield  {author} {\bibinfo {author} {\bibfnamefont {J.}~\bibnamefont
  {Elgeti}}, \bibinfo {author} {\bibfnamefont {R.~G.}\ \bibnamefont
  {Winkler}},\ and\ \bibinfo {author} {\bibfnamefont {G.}~\bibnamefont
  {Gompper}},\ }\bibfield  {title} {\enquote {\bibinfo {title} {Physics of
  microswimmer--single particle motion and collective behavior: a review},}\
  }\href {https://doi.org/10.1088/0034-4885/78/5/056601} {\bibfield  {journal}
  {\bibinfo  {journal} {Rep. Prog. Phys.}\ }\textbf {\bibinfo {volume} {78}},\
  \bibinfo {pages} {056601} (\bibinfo {year} {2015})}\BibitemShut {NoStop}%
\bibitem [{\citenamefont {Bechinger}\ \emph {et~al.}(2016)\citenamefont
  {Bechinger}, \citenamefont {Di~Leonardo}, \citenamefont {L\"owen},
  \citenamefont {Reichhardt}, \citenamefont {Volpe},\ and\ \citenamefont
  {Volpe}}]{bechinger2016ap}%
  \BibitemOpen
  \bibfield  {author} {\bibinfo {author} {\bibfnamefont {C.}~\bibnamefont
  {Bechinger}}, \bibinfo {author} {\bibfnamefont {R.}~\bibnamefont
  {Di~Leonardo}}, \bibinfo {author} {\bibfnamefont {H.}~\bibnamefont
  {L\"owen}}, \bibinfo {author} {\bibfnamefont {C.}~\bibnamefont {Reichhardt}},
  \bibinfo {author} {\bibfnamefont {G.}~\bibnamefont {Volpe}},\ and\ \bibinfo
  {author} {\bibfnamefont {G.}~\bibnamefont {Volpe}},\ }\bibfield  {title}
  {\enquote {\bibinfo {title} {Active particles in complex and crowded
  environments},}\ }\href {https://doi.org/10.1103/RevModPhys.88.045006}
  {\bibfield  {journal} {\bibinfo  {journal} {Rev. Mod. Phys.}\ }\textbf
  {\bibinfo {volume} {88}},\ \bibinfo {pages} {045006} (\bibinfo {year}
  {2016})}\BibitemShut {NoStop}%
\bibitem [{\citenamefont {Paxton}\ \emph {et~al.}(2004)\citenamefont {Paxton},
  \citenamefont {Kistler}, \citenamefont {Olmeda}, \citenamefont {Sen},
  \citenamefont {St.~Angelo}, \citenamefont {Cao}, \citenamefont {Mallouk},
  \citenamefont {Lammert},\ and\ \citenamefont {Crespi}}]{paxton2004catalytic}%
  \BibitemOpen
  \bibfield  {author} {\bibinfo {author} {\bibfnamefont {W.~F.}\ \bibnamefont
  {Paxton}}, \bibinfo {author} {\bibfnamefont {K.~C.}\ \bibnamefont {Kistler}},
  \bibinfo {author} {\bibfnamefont {C.~C.}\ \bibnamefont {Olmeda}}, \bibinfo
  {author} {\bibfnamefont {A.}~\bibnamefont {Sen}}, \bibinfo {author}
  {\bibfnamefont {S.~K.}\ \bibnamefont {St.~Angelo}}, \bibinfo {author}
  {\bibfnamefont {Y.}~\bibnamefont {Cao}}, \bibinfo {author} {\bibfnamefont
  {T.~E.}\ \bibnamefont {Mallouk}}, \bibinfo {author} {\bibfnamefont {P.~E.}\
  \bibnamefont {Lammert}},\ and\ \bibinfo {author} {\bibfnamefont {V.~H.}\
  \bibnamefont {Crespi}},\ }\bibfield  {title} {\enquote {\bibinfo {title}
  {Catalytic nanomotors: Autonomous movement of striped nanorods},}\ }\href
  {https://doi.org/10.1021/ja047697z} {\bibfield  {journal} {\bibinfo
  {journal} {J. Am. Chem. Soc.}\ }\textbf {\bibinfo {volume} {126}},\ \bibinfo
  {pages} {13424--13431} (\bibinfo {year} {2004})}\BibitemShut {NoStop}%
\bibitem [{\citenamefont {Vicsek}\ and\ \citenamefont
  {Zafeiris}(2010)}]{vicsek2010coll}%
  \BibitemOpen
  \bibfield  {author} {\bibinfo {author} {\bibfnamefont {T.}~\bibnamefont
  {Vicsek}}\ and\ \bibinfo {author} {\bibfnamefont {A.}~\bibnamefont
  {Zafeiris}},\ }\bibfield  {title} {\enquote {\bibinfo {title} {Collective
  motion},}\ }\href {https://doi.org/10.1016/j.physrep.2012.03.004} {\bibfield
  {journal} {\bibinfo  {journal} {Phys. Rep.}\ }\textbf {\bibinfo {volume}
  {517}} (\bibinfo {year} {2010})}\BibitemShut {NoStop}%
\bibitem [{\citenamefont {Cates}\ and\ \citenamefont
  {Tailleur}(2014)}]{cates2014mips}%
  \BibitemOpen
  \bibfield  {author} {\bibinfo {author} {\bibfnamefont {M.}~\bibnamefont
  {Cates}}\ and\ \bibinfo {author} {\bibfnamefont {J.}~\bibnamefont
  {Tailleur}},\ }\bibfield  {title} {\enquote {\bibinfo {title}
  {Motility-induced phase separation},}\ }\href
  {https://doi.org/10.1146/annurev-conmatphys-031214-014710} {\bibfield
  {journal} {\bibinfo  {journal} {Annu. Rev. Condens. Matter Phys.}\ }\textbf
  {\bibinfo {volume} {6}} (\bibinfo {year} {2014})}\BibitemShut {NoStop}%
\bibitem [{\citenamefont {Solon}\ and\ \citenamefont
  {et~al.}(2015)}]{solon2015pressure}%
  \BibitemOpen
  \bibfield  {author} {\bibinfo {author} {\bibfnamefont {A.~P.}\ \bibnamefont
  {Solon}}\ and\ \bibinfo {author} {\bibnamefont {et~al.}},\ }\bibfield
  {title} {\enquote {\bibinfo {title} {Pressure is not a state function for
  generic active fluids},}\ }\href {https://doi.org/10.1038/nphys3377}
  {\bibfield  {journal} {\bibinfo  {journal} {Nature Physics}\ }\textbf
  {\bibinfo {volume} {11}},\ \bibinfo {pages} {673--678} (\bibinfo {year}
  {2015})}\BibitemShut {NoStop}%
\bibitem [{\citenamefont {Chat\'e}\ \emph {et~al.}(2008)\citenamefont
  {Chat\'e}, \citenamefont {Ginelli}, \citenamefont {Gr\'egoire},\ and\
  \citenamefont {Raynaud}}]{chat2008coll}%
  \BibitemOpen
  \bibfield  {author} {\bibinfo {author} {\bibfnamefont {H.}~\bibnamefont
  {Chat\'e}}, \bibinfo {author} {\bibfnamefont {F.}~\bibnamefont {Ginelli}},
  \bibinfo {author} {\bibfnamefont {G.}~\bibnamefont {Gr\'egoire}},\ and\
  \bibinfo {author} {\bibfnamefont {F.}~\bibnamefont {Raynaud}},\ }\bibfield
  {title} {\enquote {\bibinfo {title} {Collective motion of self-propelled
  particles interacting without cohesion},}\ }\href
  {https://doi.org/10.1103/PhysRevE.77.046113} {\bibfield  {journal} {\bibinfo
  {journal} {Phys. Rev. E}\ }\textbf {\bibinfo {volume} {77}},\ \bibinfo
  {pages} {046113} (\bibinfo {year} {2008})}\BibitemShut {NoStop}%
\bibitem [{\citenamefont {Ginelli}\ \emph {et~al.}(2010)\citenamefont
  {Ginelli}, \citenamefont {Peruani}, \citenamefont {B\"ar},\ and\
  \citenamefont {Chat\'e}}]{ginelli2010large}%
  \BibitemOpen
  \bibfield  {author} {\bibinfo {author} {\bibfnamefont {F.}~\bibnamefont
  {Ginelli}}, \bibinfo {author} {\bibfnamefont {F.}~\bibnamefont {Peruani}},
  \bibinfo {author} {\bibfnamefont {M.}~\bibnamefont {B\"ar}},\ and\ \bibinfo
  {author} {\bibfnamefont {H.}~\bibnamefont {Chat\'e}},\ }\bibfield  {title}
  {\enquote {\bibinfo {title} {Large-scale collective properties of
  self-propelled rods},}\ }\href
  {https://doi.org/10.1103/PhysRevLett.104.184502} {\bibfield  {journal}
  {\bibinfo  {journal} {Phys. Rev. Lett.}\ }\textbf {\bibinfo {volume} {104}},\
  \bibinfo {pages} {184502} (\bibinfo {year} {2010})}\BibitemShut {NoStop}%
\bibitem [{\citenamefont {Z{\"o}ttl}\ and\ \citenamefont
  {Stark}(2016)}]{zottl2016emergent}%
  \BibitemOpen
  \bibfield  {author} {\bibinfo {author} {\bibfnamefont {A.}~\bibnamefont
  {Z{\"o}ttl}}\ and\ \bibinfo {author} {\bibfnamefont {H.}~\bibnamefont
  {Stark}},\ }\bibfield  {title} {\enquote {\bibinfo {title} {Emergent behavior
  in active colloids},}\ }\href
  {https://doi.org/article/10.1088/0953-8984/28/25/253001} {\bibfield
  {journal} {\bibinfo  {journal} {J. Condens. Matter Phys.}\ }\textbf {\bibinfo
  {volume} {28}},\ \bibinfo {pages} {253001} (\bibinfo {year}
  {2016})}\BibitemShut {NoStop}%
\bibitem [{\citenamefont {Lauga}(2016)}]{lauga2016bacterial}%
  \BibitemOpen
  \bibfield  {author} {\bibinfo {author} {\bibfnamefont {E.}~\bibnamefont
  {Lauga}},\ }\bibfield  {title} {\enquote {\bibinfo {title} {Bacterial
  hydrodynamics},}\ }\href
  {https://doi.org/10.1146/annurev-fluid-122414-034606} {\bibfield  {journal}
  {\bibinfo  {journal} {Annu. Rev. Fluid Mech.}\ }\textbf {\bibinfo {volume}
  {48}},\ \bibinfo {pages} {105--130} (\bibinfo {year} {2016})}\BibitemShut
  {NoStop}%
\bibitem [{\citenamefont {Toner}, \citenamefont {Tu},\ and\ \citenamefont
  {Ramaswamy}(2005)}]{toner2005hydrodynamics}%
  \BibitemOpen
  \bibfield  {author} {\bibinfo {author} {\bibfnamefont {J.}~\bibnamefont
  {Toner}}, \bibinfo {author} {\bibfnamefont {Y.}~\bibnamefont {Tu}},\ and\
  \bibinfo {author} {\bibfnamefont {S.}~\bibnamefont {Ramaswamy}},\ }\bibfield
  {title} {\enquote {\bibinfo {title} {Hydrodynamics and phases of flocks},}\
  }\href {https://doi.org/10.1016/j.aop.2005.04.011} {\bibfield  {journal}
  {\bibinfo  {journal} {Ann Phys (N Y)}\ }\textbf {\bibinfo {volume} {318}},\
  \bibinfo {pages} {170--244} (\bibinfo {year} {2005})}\BibitemShut {NoStop}%
\bibitem [{\citenamefont {Buttinoni}\ \emph {et~al.}(2013)\citenamefont
  {Buttinoni}, \citenamefont {Bialk{\'e}}, \citenamefont {K{\"u}mmel},
  \citenamefont {L{\"o}wen}, \citenamefont {Bechinger},\ and\ \citenamefont
  {Speck}}]{buttinoni2013dynamical}%
  \BibitemOpen
  \bibfield  {author} {\bibinfo {author} {\bibfnamefont {I.}~\bibnamefont
  {Buttinoni}}, \bibinfo {author} {\bibfnamefont {J.}~\bibnamefont
  {Bialk{\'e}}}, \bibinfo {author} {\bibfnamefont {F.}~\bibnamefont
  {K{\"u}mmel}}, \bibinfo {author} {\bibfnamefont {H.}~\bibnamefont
  {L{\"o}wen}}, \bibinfo {author} {\bibfnamefont {C.}~\bibnamefont
  {Bechinger}},\ and\ \bibinfo {author} {\bibfnamefont {T.}~\bibnamefont
  {Speck}},\ }\bibfield  {title} {\enquote {\bibinfo {title} {Dynamical
  clustering and phase separation in suspensions of self-propelled colloidal
  particles},}\ }\href {https://doi.org/10.1103/PhysRevLett.110.238301}
  {\bibfield  {journal} {\bibinfo  {journal} {Phys. Rev. Lett.}\ }\textbf
  {\bibinfo {volume} {110}},\ \bibinfo {pages} {238301} (\bibinfo {year}
  {2013})}\BibitemShut {NoStop}%
\bibitem [{\citenamefont {Z{\"o}ttl}\ and\ \citenamefont
  {Stark}(2012)}]{zottl2012nonlinear}%
  \BibitemOpen
  \bibfield  {author} {\bibinfo {author} {\bibfnamefont {A.}~\bibnamefont
  {Z{\"o}ttl}}\ and\ \bibinfo {author} {\bibfnamefont {H.}~\bibnamefont
  {Stark}},\ }\bibfield  {title} {\enquote {\bibinfo {title} {Nonlinear
  dynamics of a microswimmer in poiseuille flow},}\ }\href
  {https://doi.org/10.1103/PhysRevLett.108.218104} {\bibfield  {journal}
  {\bibinfo  {journal} {Phys. Rev. Lett.}\ }\textbf {\bibinfo {volume} {108}},\
  \bibinfo {pages} {218104} (\bibinfo {year} {2012})}\BibitemShut {NoStop}%
\bibitem [{\citenamefont {Son}, \citenamefont {Brumley},\ and\ \citenamefont
  {Stocker}(2015{\natexlab{a}})}]{son2015live}%
  \BibitemOpen
  \bibfield  {author} {\bibinfo {author} {\bibfnamefont {K.}~\bibnamefont
  {Son}}, \bibinfo {author} {\bibfnamefont {D.~R.}\ \bibnamefont {Brumley}},\
  and\ \bibinfo {author} {\bibfnamefont {R.}~\bibnamefont {Stocker}},\
  }\bibfield  {title} {\enquote {\bibinfo {title} {Live from under the lens:
  exploring microbial motility with dynamic imaging and microfluidics},}\
  }\href {https://doi.org/10.1038/nrmicro3567} {\bibfield  {journal} {\bibinfo
  {journal} {Nat. Rev. Microbiol.}\ }\textbf {\bibinfo {volume} {13}},\
  \bibinfo {pages} {761--775} (\bibinfo {year}
  {2015}{\natexlab{a}})}\BibitemShut {NoStop}%
\bibitem [{\citenamefont {Yan}\ \emph {et~al.}(2016)\citenamefont {Yan},
  \citenamefont {Bloom}, \citenamefont {Bae}, \citenamefont {Luijten},\ and\
  \citenamefont {Granick}}]{yan2016reconfiguring}%
  \BibitemOpen
  \bibfield  {author} {\bibinfo {author} {\bibfnamefont {J.}~\bibnamefont
  {Yan}}, \bibinfo {author} {\bibfnamefont {M.}~\bibnamefont {Bloom}}, \bibinfo
  {author} {\bibfnamefont {S.~C.}\ \bibnamefont {Bae}}, \bibinfo {author}
  {\bibfnamefont {E.}~\bibnamefont {Luijten}},\ and\ \bibinfo {author}
  {\bibfnamefont {S.}~\bibnamefont {Granick}},\ }\bibfield  {title} {\enquote
  {\bibinfo {title} {Reconfiguring active particles by electrostatic
  imbalance},}\ }\href {https://doi.org/10.1038/nmat4696} {\bibfield  {journal}
  {\bibinfo  {journal} {Nat. Mater}\ }\textbf {\bibinfo {volume} {15}},\
  \bibinfo {pages} {1095--1099} (\bibinfo {year} {2016})}\BibitemShut {NoStop}%
\bibitem [{\citenamefont {Redner}, \citenamefont {Hagan},\ and\ \citenamefont
  {Baskaran}(2013)}]{redner2013structure}%
  \BibitemOpen
  \bibfield  {author} {\bibinfo {author} {\bibfnamefont {G.~S.}\ \bibnamefont
  {Redner}}, \bibinfo {author} {\bibfnamefont {M.~F.}\ \bibnamefont {Hagan}},\
  and\ \bibinfo {author} {\bibfnamefont {A.}~\bibnamefont {Baskaran}},\
  }\bibfield  {title} {\enquote {\bibinfo {title} {Structure and dynamics of a
  phase-separating active colloidal fluid},}\ }\href
  {https://doi.org/10.1103/PhysRevLett.110.055701} {\bibfield  {journal}
  {\bibinfo  {journal} {Phys. Rev. Lett.}\ }\textbf {\bibinfo {volume} {110}},\
  \bibinfo {pages} {055701} (\bibinfo {year} {2013})}\BibitemShut {NoStop}%
\bibitem [{\citenamefont {Theurkauff}\ \emph {et~al.}(2012)\citenamefont
  {Theurkauff}, \citenamefont {Cottin-Bizonne}, \citenamefont {Palacci},
  \citenamefont {Ybert},\ and\ \citenamefont
  {Bocquet}}]{theurkauff2012dynamic}%
  \BibitemOpen
  \bibfield  {author} {\bibinfo {author} {\bibfnamefont {I.}~\bibnamefont
  {Theurkauff}}, \bibinfo {author} {\bibfnamefont {C.}~\bibnamefont
  {Cottin-Bizonne}}, \bibinfo {author} {\bibfnamefont {J.}~\bibnamefont
  {Palacci}}, \bibinfo {author} {\bibfnamefont {C.}~\bibnamefont {Ybert}},\
  and\ \bibinfo {author} {\bibfnamefont {L.}~\bibnamefont {Bocquet}},\
  }\bibfield  {title} {\enquote {\bibinfo {title} {Dynamic clustering in active
  colloidal suspensions with chemical signaling},}\ }\href
  {https://doi.org/10.1103/PhysRevLett.108.268303} {\bibfield  {journal}
  {\bibinfo  {journal} {Phys. Rev. Lett.}\ }\textbf {\bibinfo {volume} {108}},\
  \bibinfo {pages} {268303} (\bibinfo {year} {2012})}\BibitemShut {NoStop}%
\bibitem [{\citenamefont {Romanczuk}\ \emph
  {et~al.}(2012{\natexlab{a}})\citenamefont {Romanczuk}, \citenamefont
  {B{\"a}r}, \citenamefont {Ebeling}, \citenamefont {Lindner},\ and\
  \citenamefont {Schimansky-Geier}}]{romanczuk2012ABP}%
  \BibitemOpen
  \bibfield  {author} {\bibinfo {author} {\bibfnamefont {P.}~\bibnamefont
  {Romanczuk}}, \bibinfo {author} {\bibfnamefont {M.}~\bibnamefont {B{\"a}r}},
  \bibinfo {author} {\bibfnamefont {W.}~\bibnamefont {Ebeling}}, \bibinfo
  {author} {\bibfnamefont {B.}~\bibnamefont {Lindner}},\ and\ \bibinfo {author}
  {\bibfnamefont {L.}~\bibnamefont {Schimansky-Geier}},\ }\bibfield  {title}
  {\enquote {\bibinfo {title} {Active brownian particles. from individual to
  collective stochastic dynamics},}\ }\href
  {https://doi.org/10.1140/epjst/e2012-01529-y} {\bibfield  {journal} {\bibinfo
   {journal} {Eur Phys J Spec Top.}\ }\textbf {\bibinfo {volume} {202}},\
  \bibinfo {pages} {1--162} (\bibinfo {year} {2012}{\natexlab{a}})}\BibitemShut
  {NoStop}%
\bibitem [{\citenamefont {Gibbons}(1982)}]{Gib1982cilia}%
  \BibitemOpen
  \bibfield  {author} {\bibinfo {author} {\bibfnamefont {I.}~\bibnamefont
  {Gibbons}},\ }\bibfield  {title} {\enquote {\bibinfo {title} {Cilia and
  flagella of eukaryotes},}\ }\href {https://doi.org/10.1083/jcb.91.3.107s}
  {\bibfield  {journal} {\bibinfo  {journal} {J. Cell Biol.}\ }\textbf
  {\bibinfo {volume} {91}},\ \bibinfo {pages} {107s--124s} (\bibinfo {year}
  {1982})}\BibitemShut {NoStop}%
\bibitem [{\citenamefont {Berg}(2003)}]{Berg2003}%
  \BibitemOpen
  \bibfield  {author} {\bibinfo {author} {\bibfnamefont {H.~C.}\ \bibnamefont
  {Berg}},\ }\bibfield  {title} {\enquote {\bibinfo {title} {The rotary motor
  of bacterial flagella},}\ }\href
  {https://doi.org/10.1146/annurev.biochem.72.121801.161737} {\bibfield
  {journal} {\bibinfo  {journal} {Annu. Rev. Biochem.}\ }\textbf {\bibinfo
  {volume} {72}},\ \bibinfo {pages} {19--54} (\bibinfo {year}
  {2003})}\BibitemShut {NoStop}%
\bibitem [{\citenamefont {Howse}\ \emph {et~al.}(2007)\citenamefont {Howse},
  \citenamefont {Jones}, \citenamefont {Ryan}, \citenamefont {Gough},
  \citenamefont {Vafabakhsh},\ and\ \citenamefont {Golestanian}}]{Howse2007}%
  \BibitemOpen
  \bibfield  {author} {\bibinfo {author} {\bibfnamefont {J.~R.}\ \bibnamefont
  {Howse}}, \bibinfo {author} {\bibfnamefont {R.~A.~L.}\ \bibnamefont {Jones}},
  \bibinfo {author} {\bibfnamefont {A.~J.}\ \bibnamefont {Ryan}}, \bibinfo
  {author} {\bibfnamefont {T.}~\bibnamefont {Gough}}, \bibinfo {author}
  {\bibfnamefont {R.}~\bibnamefont {Vafabakhsh}},\ and\ \bibinfo {author}
  {\bibfnamefont {R.}~\bibnamefont {Golestanian}},\ }\bibfield  {title}
  {\enquote {\bibinfo {title} {Self-motile colloidal particles: From directed
  propulsion to random walk},}\ }\href
  {https://doi.org/10.1103/PhysRevLett.99.048102} {\bibfield  {journal}
  {\bibinfo  {journal} {Phys. Rev. Lett.}\ }\textbf {\bibinfo {volume} {99}},\
  \bibinfo {pages} {048102} (\bibinfo {year} {2007})}\BibitemShut {NoStop}%
\bibitem [{\citenamefont {Rao}\ \emph {et~al.}(2018)\citenamefont {Rao},
  \citenamefont {Reddy}, \citenamefont {Fransaer},\ and\ \citenamefont
  {Clasen}}]{rao2018self}%
  \BibitemOpen
  \bibfield  {author} {\bibinfo {author} {\bibfnamefont {D.~V.}\ \bibnamefont
  {Rao}}, \bibinfo {author} {\bibfnamefont {N.}~\bibnamefont {Reddy}}, \bibinfo
  {author} {\bibfnamefont {J.}~\bibnamefont {Fransaer}},\ and\ \bibinfo
  {author} {\bibfnamefont {C.}~\bibnamefont {Clasen}},\ }\bibfield  {title}
  {\enquote {\bibinfo {title} {Self-propulsion of bent bimetallic janus
  rods},}\ }\href {https://doi.org/10.1088/1361-6463/aae6f6} {\bibfield
  {journal} {\bibinfo  {journal} {J. Phys. D: Appl. Phys.}\ }\textbf {\bibinfo
  {volume} {52}},\ \bibinfo {pages} {014002} (\bibinfo {year}
  {2018})}\BibitemShut {NoStop}%
\bibitem [{\citenamefont {Poggi}\ and\ \citenamefont
  {Gohy}(2017)}]{poggi2017janus}%
  \BibitemOpen
  \bibfield  {author} {\bibinfo {author} {\bibfnamefont {E.}~\bibnamefont
  {Poggi}}\ and\ \bibinfo {author} {\bibfnamefont {J.-F.}\ \bibnamefont
  {Gohy}},\ }\bibfield  {title} {\enquote {\bibinfo {title} {Janus particles:
  from synthesis to application},}\ }\href
  {https://doi.org/10.1007/s00396-017-4192-8} {\bibfield  {journal} {\bibinfo
  {journal} {Colloid Polym. Sci.}\ }\textbf {\bibinfo {volume} {295}},\
  \bibinfo {pages} {2083--2108} (\bibinfo {year} {2017})}\BibitemShut {NoStop}%
\bibitem [{\citenamefont {B{\"a}r}\ \emph {et~al.}(2020)\citenamefont
  {B{\"a}r}, \citenamefont {Gro{\ss}mann}, \citenamefont {Heidenreich},\ and\
  \citenamefont {Peruani}}]{bar2020self}%
  \BibitemOpen
  \bibfield  {author} {\bibinfo {author} {\bibfnamefont {M.}~\bibnamefont
  {B{\"a}r}}, \bibinfo {author} {\bibfnamefont {R.}~\bibnamefont
  {Gro{\ss}mann}}, \bibinfo {author} {\bibfnamefont {S.}~\bibnamefont
  {Heidenreich}},\ and\ \bibinfo {author} {\bibfnamefont {F.}~\bibnamefont
  {Peruani}},\ }\bibfield  {title} {\enquote {\bibinfo {title} {Self-propelled
  rods: Insights and perspectives for active matter},}\ }\href
  {https://doi.org/10.1146/annurev-conmatphys-031119-050611} {\bibfield
  {journal} {\bibinfo  {journal} {Annu. Rev. Condens. Matter Phys.}\ }\textbf
  {\bibinfo {volume} {11}},\ \bibinfo {pages} {441--466} (\bibinfo {year}
  {2020})}\BibitemShut {NoStop}%
\bibitem [{\citenamefont {Berg}\ and\ \citenamefont
  {Purcell}(1977)}]{berg1977escherichia}%
  \BibitemOpen
  \bibfield  {author} {\bibinfo {author} {\bibfnamefont {H.~C.}\ \bibnamefont
  {Berg}}\ and\ \bibinfo {author} {\bibfnamefont {E.~M.}\ \bibnamefont
  {Purcell}},\ }\bibfield  {title} {\enquote {\bibinfo {title} {Physics of
  chemoreception},}\ }\href {https://doi.org/10.1016/S0006-3495(77)85544-6}
  {\bibfield  {journal} {\bibinfo  {journal} {Biophysical journal}\ }\textbf
  {\bibinfo {volume} {20}},\ \bibinfo {pages} {193--219} (\bibinfo {year}
  {1977})}\BibitemShut {NoStop}%
\bibitem [{\citenamefont {Palacci}\ \emph {et~al.}(2013)\citenamefont
  {Palacci}, \citenamefont {Sacanna}, \citenamefont {Steinberg}, \citenamefont
  {Pine},\ and\ \citenamefont {Chaikin}}]{palacci2013living}%
  \BibitemOpen
  \bibfield  {author} {\bibinfo {author} {\bibfnamefont {J.}~\bibnamefont
  {Palacci}}, \bibinfo {author} {\bibfnamefont {S.}~\bibnamefont {Sacanna}},
  \bibinfo {author} {\bibfnamefont {A.~P.}\ \bibnamefont {Steinberg}}, \bibinfo
  {author} {\bibfnamefont {D.~J.}\ \bibnamefont {Pine}},\ and\ \bibinfo
  {author} {\bibfnamefont {P.~M.}\ \bibnamefont {Chaikin}},\ }\bibfield
  {title} {\enquote {\bibinfo {title} {Living crystals of light-activated
  colloidal surfers},}\ }\href {https://doi.org/10.1126/science.1230020}
  {\bibfield  {journal} {\bibinfo  {journal} {Science}\ }\textbf {\bibinfo
  {volume} {339}},\ \bibinfo {pages} {936--940} (\bibinfo {year}
  {2013})}\BibitemShut {NoStop}%
\bibitem [{\citenamefont {Goldstein}(2015)}]{goldstein2015green}%
  \BibitemOpen
  \bibfield  {author} {\bibinfo {author} {\bibfnamefont {R.~E.}\ \bibnamefont
  {Goldstein}},\ }\bibfield  {title} {\enquote {\bibinfo {title} {Green algae
  as model organisms for biological fluid dynamics},}\ }\href
  {https://doi.org/10.1146/annurev-fluid-010313-141426} {\bibfield  {journal}
  {\bibinfo  {journal} {Annu. Rev. Fluid Mech.}\ }\textbf {\bibinfo {volume}
  {47}},\ \bibinfo {pages} {343--375} (\bibinfo {year} {2015})}\BibitemShut
  {NoStop}%
\bibitem [{\citenamefont {Blakemore}(1975)}]{blakemore1975magnetotactic}%
  \BibitemOpen
  \bibfield  {author} {\bibinfo {author} {\bibfnamefont {R.~P.}\ \bibnamefont
  {Blakemore}},\ }\bibfield  {title} {\enquote {\bibinfo {title} {Magnetotactic
  bacteria},}\ }\href {https://doi.org/10.1126/science.170679} {\bibfield
  {journal} {\bibinfo  {journal} {Science (New York, N.Y.)}\ }\textbf {\bibinfo
  {volume} {190}},\ \bibinfo {pages} {377--379} (\bibinfo {year}
  {1975})}\BibitemShut {NoStop}%
\bibitem [{\citenamefont {Erglis}\ \emph {et~al.}(2007)\citenamefont {Erglis},
  \citenamefont {Wen}, \citenamefont {Ose}, \citenamefont {Zeltins},
  \citenamefont {Sharipo}, \citenamefont {Janmey},\ and\ \citenamefont
  {Cebers}}]{erglis2007mag}%
  \BibitemOpen
  \bibfield  {author} {\bibinfo {author} {\bibfnamefont {K.}~\bibnamefont
  {Erglis}}, \bibinfo {author} {\bibfnamefont {Q.}~\bibnamefont {Wen}},
  \bibinfo {author} {\bibfnamefont {V.}~\bibnamefont {Ose}}, \bibinfo {author}
  {\bibfnamefont {A.}~\bibnamefont {Zeltins}}, \bibinfo {author} {\bibfnamefont
  {A.}~\bibnamefont {Sharipo}}, \bibinfo {author} {\bibfnamefont
  {P.}~\bibnamefont {Janmey}},\ and\ \bibinfo {author} {\bibfnamefont
  {A.}~\bibnamefont {Cebers}},\ }\bibfield  {title} {\enquote {\bibinfo {title}
  {Dynamics of magnetotactic bacteria in a rotating magnetic field},}\ }\href
  {https://doi.org/10.1529/biophysj.107.107474} {\bibfield  {journal} {\bibinfo
   {journal} {Biophysical journal}\ }\textbf {\bibinfo {volume} {93}},\
  \bibinfo {pages} {1402--12} (\bibinfo {year} {2007})}\BibitemShut {NoStop}%
\bibitem [{\citenamefont {Shabanniya}\ and\ \citenamefont
  {Naji}(2020)}]{shabanniya2020active}%
  \BibitemOpen
  \bibfield  {author} {\bibinfo {author} {\bibfnamefont {M.~R.}\ \bibnamefont
  {Shabanniya}}\ and\ \bibinfo {author} {\bibfnamefont {A.}~\bibnamefont
  {Naji}},\ }\bibfield  {title} {\enquote {\bibinfo {title} {Active dipolar
  spheroids in shear flow and transverse field: Population splitting,
  cross-stream migration, and orientational pinning},}\ }\href
  {https://doi.org/10.1063/5.0002757} {\bibfield  {journal} {\bibinfo
  {journal} {J. Chem. Phys.}\ }\textbf {\bibinfo {volume} {152}} (\bibinfo
  {year} {2020})}\BibitemShut {NoStop}%
\bibitem [{\citenamefont {Marchetti}\ \emph {et~al.}(2016)\citenamefont
  {Marchetti}, \citenamefont {Fily}, \citenamefont {Henkes}, \citenamefont
  {Patch},\ and\ \citenamefont {Yllanes}}]{MARCHETTI201634}%
  \BibitemOpen
  \bibfield  {author} {\bibinfo {author} {\bibfnamefont {M.~C.}\ \bibnamefont
  {Marchetti}}, \bibinfo {author} {\bibfnamefont {Y.}~\bibnamefont {Fily}},
  \bibinfo {author} {\bibfnamefont {S.}~\bibnamefont {Henkes}}, \bibinfo
  {author} {\bibfnamefont {A.}~\bibnamefont {Patch}},\ and\ \bibinfo {author}
  {\bibfnamefont {D.}~\bibnamefont {Yllanes}},\ }\bibfield  {title} {\enquote
  {\bibinfo {title} {Minimal model of active colloids highlights the role of
  mechanical interactions in controlling the emergent behavior of active
  matter},}\ }\href
  {https://doi.org/https://doi.org/10.1016/j.cocis.2016.01.003} {\bibfield
  {journal} {\bibinfo  {journal} {Current Opinion in Colloid \& Interface
  Science}\ }\textbf {\bibinfo {volume} {21}},\ \bibinfo {pages} {34--43}
  (\bibinfo {year} {2016})}\BibitemShut {NoStop}%
\bibitem [{\citenamefont {Schneider}\ and\ \citenamefont
  {Stark}(2019)}]{stark2019opt}%
  \BibitemOpen
  \bibfield  {author} {\bibinfo {author} {\bibfnamefont {E.}~\bibnamefont
  {Schneider}}\ and\ \bibinfo {author} {\bibfnamefont {H.}~\bibnamefont
  {Stark}},\ }\bibfield  {title} {\enquote {\bibinfo {title} {Optimal steering
  of a smart active particle},}\ }\href
  {https://doi.org/10.1209/0295-5075/127/64003} {\bibfield  {journal} {\bibinfo
   {journal} {EPL}\ }\textbf {\bibinfo {volume} {127}},\ \bibinfo {pages}
  {64003} (\bibinfo {year} {2019})}\BibitemShut {NoStop}%
\bibitem [{\citenamefont {Colabrese}\ \emph {et~al.}(2017)\citenamefont
  {Colabrese}, \citenamefont {Gustavsson}, \citenamefont {Celani},\ and\
  \citenamefont {Biferale}}]{colabrese2017flow}%
  \BibitemOpen
  \bibfield  {author} {\bibinfo {author} {\bibfnamefont {S.}~\bibnamefont
  {Colabrese}}, \bibinfo {author} {\bibfnamefont {K.}~\bibnamefont
  {Gustavsson}}, \bibinfo {author} {\bibfnamefont {A.}~\bibnamefont {Celani}},\
  and\ \bibinfo {author} {\bibfnamefont {L.}~\bibnamefont {Biferale}},\
  }\bibfield  {title} {\enquote {\bibinfo {title} {Flow navigation by smart
  microswimmers via reinforcement learning},}\ }\href
  {https://doi.org/10.1103/PhysRevLett.118.158004} {\bibfield  {journal}
  {\bibinfo  {journal} {Phys. Rev. Lett.}\ }\textbf {\bibinfo {volume} {118}},\
  \bibinfo {pages} {158004} (\bibinfo {year} {2017})}\BibitemShut {NoStop}%
\bibitem [{\citenamefont {Gustavsson}\ \emph {et~al.}(2017)\citenamefont
  {Gustavsson}, \citenamefont {Biferale}, \citenamefont {Celani},\ and\
  \citenamefont {Colabrese}}]{Gustavsson2017}%
  \BibitemOpen
  \bibfield  {author} {\bibinfo {author} {\bibfnamefont {K.}~\bibnamefont
  {Gustavsson}}, \bibinfo {author} {\bibfnamefont {L.}~\bibnamefont
  {Biferale}}, \bibinfo {author} {\bibfnamefont {A.}~\bibnamefont {Celani}},\
  and\ \bibinfo {author} {\bibfnamefont {S.}~\bibnamefont {Colabrese}},\
  }\bibfield  {title} {\enquote {\bibinfo {title} {Finding efficient swimming
  strategies in a three-dimensional chaotic flow by reinforcement learning},}\
  }\href {https://doi.org/10.1140/epje/i2017-11602-9} {\bibfield  {journal}
  {\bibinfo  {journal} {Eur Phys J E Soft Matter EUR PHYS J E}\ }\textbf
  {\bibinfo {volume} {40}},\ \bibinfo {pages} {110} (\bibinfo {year}
  {2017})}\BibitemShut {NoStop}%
\bibitem [{\citenamefont {Biferale}\ \emph {et~al.}(2019)\citenamefont
  {Biferale}, \citenamefont {Bonaccorso}, \citenamefont {Buzzicotti},
  \citenamefont {Clark Di~Leoni},\ and\ \citenamefont
  {Gustavsson}}]{biferale2019zermelo}%
  \BibitemOpen
  \bibfield  {author} {\bibinfo {author} {\bibfnamefont {L.}~\bibnamefont
  {Biferale}}, \bibinfo {author} {\bibfnamefont {F.}~\bibnamefont
  {Bonaccorso}}, \bibinfo {author} {\bibfnamefont {M.}~\bibnamefont
  {Buzzicotti}}, \bibinfo {author} {\bibfnamefont {P.}~\bibnamefont {Clark
  Di~Leoni}},\ and\ \bibinfo {author} {\bibfnamefont {K.}~\bibnamefont
  {Gustavsson}},\ }\bibfield  {title} {\enquote {\bibinfo {title} {Zermelo's
  problem: optimal point-to-point navigation in 2d turbulent flows using
  reinforcement learning},}\ }\href {https://doi.org/10.1063/1.5120370}
  {\bibfield  {journal} {\bibinfo  {journal} {Chaos}\ }\textbf {\bibinfo
  {volume} {29}} (\bibinfo {year} {2019})}\BibitemShut {NoStop}%
\bibitem [{\citenamefont {Nasiri}\ and\ \citenamefont
  {Liebchen}(2022)}]{nasiri2022reinforcement}%
  \BibitemOpen
  \bibfield  {author} {\bibinfo {author} {\bibfnamefont {M.}~\bibnamefont
  {Nasiri}}\ and\ \bibinfo {author} {\bibfnamefont {B.}~\bibnamefont
  {Liebchen}},\ }\bibfield  {title} {\enquote {\bibinfo {title} {Reinforcement
  learning of optimal active particle navigation},}\ }\href
  {https://doi.org/10.1088/1367-2630/ac8013} {\bibfield  {journal} {\bibinfo
  {journal} {New J. Phys.}\ }\textbf {\bibinfo {volume} {24}},\ \bibinfo
  {pages} {073042} (\bibinfo {year} {2022})}\BibitemShut {NoStop}%
\bibitem [{\citenamefont {Gunnarson}\ \emph
  {et~al.}(2021{\natexlab{a}})\citenamefont {Gunnarson}, \citenamefont
  {Mandralis}, \citenamefont {Novati}, \citenamefont {Koumoutsakos},\ and\
  \citenamefont {Dabiri}}]{gunnarson2021learn}%
  \BibitemOpen
  \bibfield  {author} {\bibinfo {author} {\bibfnamefont {P.}~\bibnamefont
  {Gunnarson}}, \bibinfo {author} {\bibfnamefont {I.}~\bibnamefont
  {Mandralis}}, \bibinfo {author} {\bibfnamefont {G.}~\bibnamefont {Novati}},
  \bibinfo {author} {\bibfnamefont {P.}~\bibnamefont {Koumoutsakos}},\ and\
  \bibinfo {author} {\bibfnamefont {J.~O.}\ \bibnamefont {Dabiri}},\ }\bibfield
   {title} {\enquote {\bibinfo {title} {Learning efficient navigation in
  vortical flow fields},}\ }\href {https://doi.org/10.1038/s41467-021-27015-y}
  {\bibfield  {journal} {\bibinfo  {journal} {Nat. Comm.}\ }\textbf {\bibinfo
  {volume} {12}},\ \bibinfo {pages} {7143} (\bibinfo {year}
  {2021}{\natexlab{a}})}\BibitemShut {NoStop}%
\bibitem [{\citenamefont {Buzzicotti}\ \emph {et~al.}(2021)\citenamefont
  {Buzzicotti}, \citenamefont {Biferale}, \citenamefont {Bonaccorso},
  \citenamefont {Clark~di Leoni},\ and\ \citenamefont
  {Gustavsson}}]{buzzicotti2021opt}%
  \BibitemOpen
  \bibfield  {author} {\bibinfo {author} {\bibfnamefont {M.}~\bibnamefont
  {Buzzicotti}}, \bibinfo {author} {\bibfnamefont {L.}~\bibnamefont
  {Biferale}}, \bibinfo {author} {\bibfnamefont {F.}~\bibnamefont
  {Bonaccorso}}, \bibinfo {author} {\bibfnamefont {P.}~\bibnamefont {Clark~di
  Leoni}},\ and\ \bibinfo {author} {\bibfnamefont {K.}~\bibnamefont
  {Gustavsson}},\ }\enquote {\bibinfo {title} {Optimal control of
  point-to-point navigation in turbulent time dependent flows using
  reinforcement learning},}\ \ (\bibinfo {year} {2021})\ pp.\ \bibinfo {pages}
  {223--234}\BibitemShut {NoStop}%
\bibitem [{\citenamefont {Zou}\ \emph {et~al.}(2022)\citenamefont {Zou},
  \citenamefont {Liu}, \citenamefont {Young}, \citenamefont {Pak},\ and\
  \citenamefont {Tsang}}]{zou2022gait}%
  \BibitemOpen
  \bibfield  {author} {\bibinfo {author} {\bibfnamefont {Z.}~\bibnamefont
  {Zou}}, \bibinfo {author} {\bibfnamefont {Y.}~\bibnamefont {Liu}}, \bibinfo
  {author} {\bibfnamefont {Y.~N.}\ \bibnamefont {Young}}, \bibinfo {author}
  {\bibfnamefont {O.~S.}\ \bibnamefont {Pak}},\ and\ \bibinfo {author}
  {\bibfnamefont {A.~C.~H.}\ \bibnamefont {Tsang}},\ }\bibfield  {title}
  {\enquote {\bibinfo {title} {Gait switching and targeted navigation of
  microswimmers via deep reinforcement learning},}\ }\href
  {https://doi.org/10.1038/s42005-022-00935-x} {\bibfield  {journal} {\bibinfo
  {journal} {Commun. Phys.}\ }\textbf {\bibinfo {volume} {5}},\ \bibinfo
  {pages} {158} (\bibinfo {year} {2022})}\BibitemShut {NoStop}%
\bibitem [{\citenamefont {Borra}\ \emph {et~al.}(2022)\citenamefont {Borra},
  \citenamefont {Biferale}, \citenamefont {Cencini},\ and\ \citenamefont
  {Celani}}]{Borra2022RL}%
  \BibitemOpen
  \bibfield  {author} {\bibinfo {author} {\bibfnamefont {F.}~\bibnamefont
  {Borra}}, \bibinfo {author} {\bibfnamefont {L.}~\bibnamefont {Biferale}},
  \bibinfo {author} {\bibfnamefont {M.}~\bibnamefont {Cencini}},\ and\ \bibinfo
  {author} {\bibfnamefont {A.}~\bibnamefont {Celani}},\ }\bibfield  {title}
  {\enquote {\bibinfo {title} {Reinforcement learning for pursuit and evasion
  of microswimmers at low reynolds number},}\ }\href
  {https://doi.org/10.1103/PhysRevFluids.7.023103} {\bibfield  {journal}
  {\bibinfo  {journal} {Phys. Rev. Fluids}\ }\textbf {\bibinfo {volume} {7}},\
  \bibinfo {pages} {023103} (\bibinfo {year} {2022})}\BibitemShut {NoStop}%
\bibitem [{\citenamefont {Mui{\~n}os-Landin}\ \emph {et~al.}(2021)\citenamefont
  {Mui{\~n}os-Landin}, \citenamefont {Fischer}, \citenamefont {Holubec},\ and\
  \citenamefont {Cichos}}]{muinos2021reinforcement}%
  \BibitemOpen
  \bibfield  {author} {\bibinfo {author} {\bibfnamefont {S.}~\bibnamefont
  {Mui{\~n}os-Landin}}, \bibinfo {author} {\bibfnamefont {A.}~\bibnamefont
  {Fischer}}, \bibinfo {author} {\bibfnamefont {V.}~\bibnamefont {Holubec}},\
  and\ \bibinfo {author} {\bibfnamefont {F.}~\bibnamefont {Cichos}},\
  }\bibfield  {title} {\enquote {\bibinfo {title} {Reinforcement learning with
  artificial microswimmers},}\ }\href
  {https://doi.org/10.1126/scirobotics.abd9285} {\bibfield  {journal} {\bibinfo
   {journal} {Sci. Robot.}\ }\textbf {\bibinfo {volume} {6}},\ \bibinfo {pages}
  {eabd9285} (\bibinfo {year} {2021})}\BibitemShut {NoStop}%
\bibitem [{\citenamefont {Putzke}\ and\ \citenamefont
  {Stark}(2023)}]{Putzke2023}%
  \BibitemOpen
  \bibfield  {author} {\bibinfo {author} {\bibfnamefont {M.}~\bibnamefont
  {Putzke}}\ and\ \bibinfo {author} {\bibfnamefont {H.}~\bibnamefont {Stark}},\
  }\bibfield  {title} {\enquote {\bibinfo {title} {Optimal navigation of a
  smart active particle: directional and distance sensing},}\ }\href
  {https://doi.org/10.1140/epje/s10189-023-00309-3} {\bibfield  {journal}
  {\bibinfo  {journal} {The European Physical Journal E}\ }\textbf {\bibinfo
  {volume} {46}},\ \bibinfo {pages} {48} (\bibinfo {year} {2023})}\BibitemShut
  {NoStop}%
\bibitem [{\citenamefont {Calascibetta}\ \emph {et~al.}(2023)\citenamefont
  {Calascibetta}, \citenamefont {Biferale}, \citenamefont {Borra},
  \citenamefont {Celani},\ and\ \citenamefont {Cencini}}]{Calascibetta2023}%
  \BibitemOpen
  \bibfield  {author} {\bibinfo {author} {\bibfnamefont {C.}~\bibnamefont
  {Calascibetta}}, \bibinfo {author} {\bibfnamefont {L.}~\bibnamefont
  {Biferale}}, \bibinfo {author} {\bibfnamefont {F.}~\bibnamefont {Borra}},
  \bibinfo {author} {\bibfnamefont {A.}~\bibnamefont {Celani}},\ and\ \bibinfo
  {author} {\bibfnamefont {M.}~\bibnamefont {Cencini}},\ }\bibfield  {title}
  {\enquote {\bibinfo {title} {Optimal tracking strategies in a turbulent
  flow},}\ }\href {https://doi.org/10.1038/s42005-023-01366-y} {\bibfield
  {journal} {\bibinfo  {journal} {Communications Physics}\ }\textbf {\bibinfo
  {volume} {6}},\ \bibinfo {pages} {256} (\bibinfo {year} {2023})}\BibitemShut
  {NoStop}%
\bibitem [{\citenamefont {Gunnarson}\ \emph
  {et~al.}(2021{\natexlab{b}})\citenamefont {Gunnarson}, \citenamefont
  {Mandralis}, \citenamefont {Novati}, \citenamefont {Koumoutsakos},\ and\
  \citenamefont {Dabiri}}]{gunnarson2021flow}%
  \BibitemOpen
  \bibfield  {author} {\bibinfo {author} {\bibfnamefont {P.}~\bibnamefont
  {Gunnarson}}, \bibinfo {author} {\bibfnamefont {I.}~\bibnamefont
  {Mandralis}}, \bibinfo {author} {\bibfnamefont {G.}~\bibnamefont {Novati}},
  \bibinfo {author} {\bibfnamefont {P.}~\bibnamefont {Koumoutsakos}},\ and\
  \bibinfo {author} {\bibfnamefont {J.~O.}\ \bibnamefont {Dabiri}},\ }\bibfield
   {title} {\enquote {\bibinfo {title} {Learning efficient navigation in
  vortical flow fields},}\ }\href {https://doi.org/10.1038/s41467-021-27015-y}
  {\bibfield  {journal} {\bibinfo  {journal} {Nat. Comm.}\ }\textbf {\bibinfo
  {volume} {12}},\ \bibinfo {pages} {7143} (\bibinfo {year}
  {2021}{\natexlab{b}})}\BibitemShut {NoStop}%
\bibitem [{\citenamefont {Son}, \citenamefont {Brumley},\ and\ \citenamefont
  {Stocker}(2015{\natexlab{b}})}]{Son2015}%
  \BibitemOpen
  \bibfield  {author} {\bibinfo {author} {\bibfnamefont {K.}~\bibnamefont
  {Son}}, \bibinfo {author} {\bibfnamefont {D.~R.}\ \bibnamefont {Brumley}},\
  and\ \bibinfo {author} {\bibfnamefont {R.}~\bibnamefont {Stocker}},\
  }\bibfield  {title} {\enquote {\bibinfo {title} {Live from under the lens:
  exploring microbial motility with dynamic imaging and microfluidics},}\
  }\href {https://doi.org/10.1038/nrmicro3567} {\bibfield  {journal} {\bibinfo
  {journal} {Nature Reviews Microbiology}\ }\textbf {\bibinfo {volume} {13}},\
  \bibinfo {pages} {761--775} (\bibinfo {year}
  {2015}{\natexlab{b}})}\BibitemShut {NoStop}%
\bibitem [{\citenamefont {Shields}, \citenamefont {Reyes},\ and\ \citenamefont
  {L{\'o}pez}(2015)}]{Shields2015}%
  \BibitemOpen
  \bibfield  {author} {\bibinfo {author} {\bibfnamefont {C.~W.}\ \bibnamefont
  {Shields}}, \bibinfo {author} {\bibfnamefont {C.~D.}\ \bibnamefont {Reyes}},\
  and\ \bibinfo {author} {\bibfnamefont {G.~P.}\ \bibnamefont {L{\'o}pez}},\
  }\bibfield  {title} {\enquote {\bibinfo {title} {Microfluidic cell sorting: a
  review of the advances in the separation of cells from debulking to rare cell
  isolation},}\ }\href {https://doi.org/10.1039/c4lc01246a} {\bibfield
  {journal} {\bibinfo  {journal} {Lab Chip}\ }\textbf {\bibinfo {volume}
  {15}},\ \bibinfo {pages} {1230--1249} (\bibinfo {year} {2015})}\BibitemShut
  {NoStop}%
\bibitem [{\citenamefont {Lenshof}\ and\ \citenamefont
  {Laurell}(2010)}]{Lenshof2010}%
  \BibitemOpen
  \bibfield  {author} {\bibinfo {author} {\bibfnamefont {A.}~\bibnamefont
  {Lenshof}}\ and\ \bibinfo {author} {\bibfnamefont {T.}~\bibnamefont
  {Laurell}},\ }\bibfield  {title} {\enquote {\bibinfo {title} {Continuous
  separation of cells and particles in microfluidic systems},}\ }\href
  {https://doi.org/https://doi.org/10.1039/B915999C} {\bibfield  {journal}
  {\bibinfo  {journal} {Chemical Society Reviews}\ }\textbf {\bibinfo {volume}
  {39}},\ \bibinfo {pages} {1203--1217} (\bibinfo {year} {2010})}\BibitemShut
  {NoStop}%
\bibitem [{\citenamefont {Sutton}\ and\ \citenamefont
  {Barto}(2018)}]{sutton2018reinforcement}%
  \BibitemOpen
  \bibfield  {author} {\bibinfo {author} {\bibfnamefont {R.~S.}\ \bibnamefont
  {Sutton}}\ and\ \bibinfo {author} {\bibfnamefont {A.~G.}\ \bibnamefont
  {Barto}},\ }\href
  {https://mitpress.mit.edu/9780262039246/reinforcement-learning/} {\emph
  {\bibinfo {title} {Reinforcement Learning: An Introduction}}}\ (\bibinfo
  {publisher} {MIT press Cambridge},\ \bibinfo {year} {2018})\BibitemShut
  {NoStop}%
\bibitem [{\citenamefont {Zhou}, \citenamefont {Huang},\ and\ \citenamefont
  {Fr{\"a}nti}(2022)}]{Rev_motion_planning}%
  \BibitemOpen
  \bibfield  {author} {\bibinfo {author} {\bibfnamefont {C.}~\bibnamefont
  {Zhou}}, \bibinfo {author} {\bibfnamefont {B.}~\bibnamefont {Huang}},\ and\
  \bibinfo {author} {\bibfnamefont {P.}~\bibnamefont {Fr{\"a}nti}},\ }\bibfield
   {title} {\enquote {\bibinfo {title} {A review of motion planning algorithms
  for intelligent robots},}\ }\href
  {https://doi.org/10.1007/s10845-021-01867-z} {\bibfield  {journal} {\bibinfo
  {journal} {J. Intell. Manuf.}\ }\textbf {\bibinfo {volume} {33}},\ \bibinfo
  {pages} {387--424} (\bibinfo {year} {2022})}\BibitemShut {NoStop}%
\bibitem [{\citenamefont {Morales}(2020)}]{morales2020grokking}%
  \BibitemOpen
  \bibfield  {author} {\bibinfo {author} {\bibfnamefont {M.}~\bibnamefont
  {Morales}},\ }\href
  {https://www.manning.com/books/grokking-deep-reinforcement-learning} {\emph
  {\bibinfo {title} {Grokking Deep Reinforcement Learning}}}\ (\bibinfo
  {publisher} {Manning Publications},\ \bibinfo {year} {2020})\BibitemShut
  {NoStop}%
\bibitem [{\citenamefont {Paszke}\ \emph {et~al.}(2019)\citenamefont {Paszke},
  \citenamefont {Gross}, \citenamefont {Massa}, \citenamefont {Lerer},
  \citenamefont {Bradbury}, \citenamefont {Chanan}, \citenamefont {Killeen},
  \citenamefont {Lin}, \citenamefont {Gimelshein}, \citenamefont {Antiga} \emph
  {et~al.}}]{paszke2019pytorch}%
  \BibitemOpen
  \bibfield  {author} {\bibinfo {author} {\bibfnamefont {A.}~\bibnamefont
  {Paszke}}, \bibinfo {author} {\bibfnamefont {S.}~\bibnamefont {Gross}},
  \bibinfo {author} {\bibfnamefont {F.}~\bibnamefont {Massa}}, \bibinfo
  {author} {\bibfnamefont {A.}~\bibnamefont {Lerer}}, \bibinfo {author}
  {\bibfnamefont {J.}~\bibnamefont {Bradbury}}, \bibinfo {author}
  {\bibfnamefont {G.}~\bibnamefont {Chanan}}, \bibinfo {author} {\bibfnamefont
  {T.}~\bibnamefont {Killeen}}, \bibinfo {author} {\bibfnamefont
  {Z.}~\bibnamefont {Lin}}, \bibinfo {author} {\bibfnamefont {N.}~\bibnamefont
  {Gimelshein}}, \bibinfo {author} {\bibfnamefont {L.}~\bibnamefont {Antiga}},
  \emph {et~al.},\ }\bibfield  {title} {\enquote {\bibinfo {title} {Pytorch: An
  imperative style, high-performance deep learning library},}\ }\href@noop {}
  {\bibfield  {journal} {\bibinfo  {journal} {Advances in Neural Information
  Processing Systems}\ }\textbf {\bibinfo {volume} {32}},\ \bibinfo {pages}
  {8024--8035} (\bibinfo {year} {2019})}\BibitemShut {NoStop}%
\bibitem [{\citenamefont {Kularatne}, \citenamefont {Bhattacharya},\ and\
  \citenamefont {Hsieh}(2018)}]{Kularatne:2018aa}%
  \BibitemOpen
  \bibfield  {author} {\bibinfo {author} {\bibfnamefont {D.}~\bibnamefont
  {Kularatne}}, \bibinfo {author} {\bibfnamefont {S.}~\bibnamefont
  {Bhattacharya}},\ and\ \bibinfo {author} {\bibfnamefont {M.~A.}\ \bibnamefont
  {Hsieh}},\ }\bibfield  {title} {\enquote {\bibinfo {title} {Going with the
  flow: a graph based approach to optimal path planning in general flows},}\
  }\href {https://doi.org/10.1007/s10514-018-9741-6} {\bibfield  {journal}
  {\bibinfo  {journal} {Autonomous Robots}\ }\textbf {\bibinfo {volume} {42}},\
  \bibinfo {pages} {1369--1387} (\bibinfo {year} {2018})}\BibitemShut {NoStop}%
\bibitem [{\citenamefont {Kim}\ and\ \citenamefont
  {Karrila}(2005)}]{kim2005microhydrodynamics}%
  \BibitemOpen
  \bibfield  {author} {\bibinfo {author} {\bibfnamefont {S.}~\bibnamefont
  {Kim}}\ and\ \bibinfo {author} {\bibfnamefont {S.~J.}\ \bibnamefont
  {Karrila}},\ }\href
  {https://www.sciencedirect.com/book/9780750691734/microhydrodynamics} {\emph
  {\bibinfo {title} {Microhydrodynamics: Principles and Selected
  Applications}}}\ (\bibinfo  {publisher} {Dover Publications},\ \bibinfo
  {year} {2005})\BibitemShut {NoStop}%
\bibitem [{\citenamefont {Happel}\ and\ \citenamefont
  {Brenner}(1983)}]{happel1983low}%
  \BibitemOpen
  \bibfield  {author} {\bibinfo {author} {\bibfnamefont {J.}~\bibnamefont
  {Happel}}\ and\ \bibinfo {author} {\bibfnamefont {H.}~\bibnamefont
  {Brenner}},\ }\href
  {https://link.springer.com/book/10.1007/978-94-009-8352-6} {\emph {\bibinfo
  {title} {Low Reynolds Number Hydrodynamics: with special applications to
  particulate media}}}\ (\bibinfo  {publisher} {Springer Netherlands \&
  Martinus Nijhoff Publishers},\ \bibinfo {address} {The Hague},\ \bibinfo
  {year} {1983})\BibitemShut {NoStop}%
\bibitem [{\citenamefont {Romanczuk}\ \emph
  {et~al.}(2012{\natexlab{b}})\citenamefont {Romanczuk}, \citenamefont {Bär},
  \citenamefont {Ebeling}, \citenamefont {Lindner},\ and\ \citenamefont
  {Schimansky-Geier}}]{romanczuk2012active}%
  \BibitemOpen
  \bibfield  {author} {\bibinfo {author} {\bibfnamefont {P.}~\bibnamefont
  {Romanczuk}}, \bibinfo {author} {\bibfnamefont {M.}~\bibnamefont {Bär}},
  \bibinfo {author} {\bibfnamefont {W.}~\bibnamefont {Ebeling}}, \bibinfo
  {author} {\bibfnamefont {B.}~\bibnamefont {Lindner}},\ and\ \bibinfo {author}
  {\bibfnamefont {L.}~\bibnamefont {Schimansky-Geier}},\ }\bibfield  {title}
  {\enquote {\bibinfo {title} {Active brownian particles: From individual to
  collective behavior},}\ }\href {https://doi.org/10.1140/epjst/e2012-01693-3}
  {\bibfield  {journal} {\bibinfo  {journal} {Eur. Phys. J.: Spec. Top.}\
  }\textbf {\bibinfo {volume} {202}},\ \bibinfo {pages} {1--162} (\bibinfo
  {year} {2012}{\natexlab{b}})}\BibitemShut {NoStop}%
\bibitem [{\citenamefont {Romanovsky}, \citenamefont {Kargovsky},\ and\
  \citenamefont {Ebeling}(2013)}]{romanovsky2013models}%
  \BibitemOpen
  \bibfield  {author} {\bibinfo {author} {\bibfnamefont {Y.}~\bibnamefont
  {Romanovsky}}, \bibinfo {author} {\bibfnamefont {A.}~\bibnamefont
  {Kargovsky}},\ and\ \bibinfo {author} {\bibfnamefont {W.}~\bibnamefont
  {Ebeling}},\ }\bibfield  {title} {\enquote {\bibinfo {title} {Models of
  active brownian motors based on internal oscillations},}\ }\href
  {https://doi.org/10.1140/epjst/e2013-01704-4} {\bibfield  {journal} {\bibinfo
   {journal} {Eur. Phys. J. Spec. Top.}\ }\textbf {\bibinfo {volume} {222}},\
  \bibinfo {pages} {2465--2474} (\bibinfo {year} {2013})}\BibitemShut {NoStop}%
\bibitem [{\citenamefont {Li}\ \emph {et~al.}(2020)\citenamefont {Li},
  \citenamefont {Li}, \citenamefont {Marchesoni}, \citenamefont {Debnath},\
  and\ \citenamefont {Ghosh}}]{li2020diffusion}%
  \BibitemOpen
  \bibfield  {author} {\bibinfo {author} {\bibfnamefont {Y.}~\bibnamefont
  {Li}}, \bibinfo {author} {\bibfnamefont {L.}~\bibnamefont {Li}}, \bibinfo
  {author} {\bibfnamefont {F.}~\bibnamefont {Marchesoni}}, \bibinfo {author}
  {\bibfnamefont {D.}~\bibnamefont {Debnath}},\ and\ \bibinfo {author}
  {\bibfnamefont {P.~K.}\ \bibnamefont {Ghosh}},\ }\bibfield  {title} {\enquote
  {\bibinfo {title} {Diffusion of chiral janus particles in convection
  rolls},}\ }\href {https://doi.org/10.1103/PhysRevResearch.2.013250}
  {\bibfield  {journal} {\bibinfo  {journal} {Physical Review Research}\
  }\textbf {\bibinfo {volume} {2}},\ \bibinfo {pages} {013250} (\bibinfo {year}
  {2020})}\BibitemShut {NoStop}%
\end{thebibliography}%

\end{document}